\documentclass[twocolumn]{aastex631}

\usepackage{amsmath}
\usepackage{xcolor}

\newcommand{\athenak}{\texttt{AthenaK}}
\newcommand{\nurates}{\texttt{bns\_nurates}}
\newcommand{\Ye}{Y_e}
\newcommand{\Pgf}{P_{\rm GF}}
\newcommand{\Mej}{M_{\rm ej}}
\newcommand{\Rns}{R_{\rm NS}}
\newcommand{\Mns}{M_{\rm NS}}
\newcommand{\gcc}{\mathrm{g\,cm^{-3}}}
\newcommand{\MeV}{\mathrm{MeV}}

\shorttitle{GR neutrino-radiation hydrodynamics of magnetar giant-flare ejecta}
\shortauthors{Pitik, Radice, Fields \& Qiu}

\begin{document}

\title{General-relativistic neutrino-radiation hydrodynamics of crust ejection in magnetar giant flares}

\author[0000-0002-9109-2451]{Tetyana Pitik}
\affiliation{Department of Physics, University of California Berkeley, Berkeley, California 94720, USA}
\email{tetyana.pitik@berkeley.edu}

\author[0000-0001-6982-1008]{David Radice}
\affiliation{Institute for Gravitation and the Cosmos, The Pennsylvania State University, University Park, PA 16802, USA}
\affiliation{Department of Physics, The Pennsylvania State University, University Park, PA 16802, USA}
\affiliation{Department of Astronomy \& Astrophysics, The Pennsylvania State University, University Park, PA 16802, USA}

\author[0000-0001-5705-1712]{Jacob Fields}
\affiliation{School of Natural Sciences, Institute for Advanced Study, 1 Einstein Drive, Princeton, NJ 08540, USA}

\author[0009-0008-0470-2460]{Yi Qiu}
\affiliation{Institute for Gravitation and the Cosmos, The Pennsylvania State University, University Park, PA 16802, USA}
\affiliation{Department of Physics, The Pennsylvania State University, University Park, PA 16802, USA}

\begin{abstract}
Magnetar giant flares are thought to eject $\gtrsim 10^{-9}\,M_\odot$ of baryons, as inferred from the radio afterglow of the 2004 flare of SGR 1806-20. One proposed mechanism is a shock driven into the neutron-star crust by the hot, high-pressure layer that the flare creates above the surface. The unbound crust could undergo $r$-process nucleosynthesis and power a short-lived "nova brevis". We study this crust ejection with general-relativistic neutrino-radiation hydrodynamics simulations performed with AthenaK, using a tabulated nuclear equation of state and two-moment neutrino transport, for flare pressures corresponding to magnetic fields $B = 10^{15}$-$10^{16.5}$ G, and compute the nucleosynthesis of the ejecta with a nuclear reaction network. We find that such flares unbind $10^{-10}$ to $5\times10^{-8}\,M_\odot$ of crust. The ejecta keep the electron fraction of the crust to within a few hundredths, and weak interactions affect it mainly in the most energetic flares. The fast, high-entropy ejecta freeze out mostly as $\alpha$ particles and free neutrons. In the stronger flares nuclei up to the third $r$-process peak do form, but those with $A \geq 90$ make up at most $\approx 8\times10^{-4}$ of the ejected mass, or $\approx 3.5\times10^{-11}\,M_\odot$ per flare, and less than $10^{-5}$ for $B \leq 10^{15.5}$ G. A nova brevis from these ejecta would be powered mainly by the decay of free neutrons. Like earlier work, our models assume spherical symmetry, a flare that covers the whole star, and no magnetic field. A field tangential to the surface would hold the crust down and weaken the shock, and a flare confined to part of the surface should eject proportionally less mass. We therefore regard our ejecta masses as upper limits for a given flare pressure, and expect realistic magnetar flares to produce negligible amounts of $r$-process material.
\end{abstract}


\section{Introduction}\label{sec:intro}

Magnetars are neutron stars (NSs) whose activity is powered by the decay of ultra-strong magnetic fields
\citep{Duncan:1992hi,Thompson:1995gw,Kouveliotou:1998ze}, see reviews by \citet{Turolla:2015mwa} and
\citet{Kaspi:2017fwg}. The most energetic of their outbursts, giant flares (GFs), release
$E \sim 10^{44}$ to a few $\times10^{46}\,$erg (isotropic equivalent) in a sub-second spike of hard X-rays and soft $\gamma$-rays followed by a
minutes-long tail modulated at the spin period. Three Galactic or Local-Group GFs have been observed, in
1979 \citep{Mazets:1979uu}, 1998 \citep{Hurley:1998ks}, and 2004 \citep{Hurley:2005zs,Palmer:2005mi}, and a growing
sample of extragalactic events has been identified in archival $\gamma$-ray data \citep{Svinkin:2021wcp,Burns:2021udk,Beniamini:2024les,Trigg:2025mim}.

The 2004 December~27 GF from SGR~1806$-$20 released $(2$--$4)\times10^{46}\,$erg for isotropic emission at
$15\,$kpc in a spike of ${\sim}0.2\,$s \citep{Hurley:2005zs,Palmer:2005mi,Cameron:2005kb,Boggs:2006uk}, followed by a
${\sim}400\,$s tail pulsating at the spin period and by a delayed, unpulsed hard component that peaked at
${\sim}600$--$800\,$s, which \citet{Mereghetti:2005dt} interpreted as an afterglow of the outflow
\citep[see also][]{Boggs:2006uk,Frederiks:2006uy}. The radio source first observed about a week later \citep{Cameron:2005kb,Gaensler:2005iq} re-brightened
after ${\sim}25$ days \citep{Gelfand:2005wi}. If the re-brightening marks
the deceleration of ejecta sweeping up the surrounding gas, the ejected mass is $\gtrsim1.6\times10^{-9}\,M_\odot$,
with an initial velocity ${\sim}0.7c$, assuming a distance of $15\,$kpc and a one-sided outflow, and up to two to three orders of magnitude larger for
plausible densities and shock microphysics \citep{Gelfand:2005wi,Granot:2005vq}. The radiation of the trapped pair
fireball ablates orders of magnitude less mass during the tail \citep{Demidov:2022jrv}, so most of the ejecta must
leave during the spike.

In the model of \citet[hereafter C24]{Cehula:2023hdh} the energy released by a rapid reconfiguration of the field fills a thin
layer of photons and $e^\pm$ pairs just above the NS surface. Its pressure $\Pgf$ can approach the magnetic pressure
of the dissipation region, $\Pgf \approx B^2/8\pi$ for a field strength $B$, and it launches a relativistic shock
into the crust. When the pressure of the layer on the crust is relieved, as the layer expands within ${\sim}10\,\mu$s, nothing holds the shock-heated outer
crust down, and its outermost part expands and becomes unbound. In one-dimensional (1D), spherically symmetric,
special-relativistic simulations, C24 obtained
$\Mej \sim (2$--$4.5)\times10^{-9}\,M_\odot\,(\Pgf/10^{30}\,{\rm erg\,cm^{-3}})^{1.43}$ with asymptotic velocities
$v \sim 0.3$--$0.6\,c$, enough for the SGR~1806$-$20 afterglow if $\Pgf \sim 10^{30}$--$10^{31}\,{\rm erg\,cm^{-3}}$.
Unlike the very neutron-rich matter decompressed from the interior of a tidally disrupted NS \citep{Lattimer:1974slx,Lattimer:1977igd},
the outer crust that is ejected has only $\Ye \approx 0.40$--$0.46$. Heavy nuclei can then form only in an
$\alpha$-rich freeze-out at high entropy and fast expansion \citep{Qian:1996xt,Hoffman:1996aj}, and C24 argued
that the hot, fast ejecta reach these conditions.

\citet{Patel:2025tse} followed the
nucleosynthesis of ejecta modeled analytically after C24 with  a nuclear reaction network, found moderate third-peak and
larger light $r$-process yields, and predicted the light curves of the radioactively powered transients
(``novae breves''). \citet{Patel:2025frn} interpreted the delayed hard component instead as the decay of freshly synthesized $r$-process nuclei in
${\sim}10^{-6}\,M_\odot$ of ejecta. Other recent work explores baryon ejection from the eruption of
magnetic loops in young magnetars \citep{Bransgrove:2025tpa}, the relativistic magnetohydrodynamics of
giant bursts and of shocks in the magnetosphere \citep{Chatterjee:2026tdp,Grehan:2026vof}, the use of novae
breves and burst spectra as probes of the crust and of its composition \citep{Zhong:2026myp,Xie:2026far}, and the
consequences of an $r$-process in the ejecta for MeV $\gamma$-ray lines \citep{Qiumu:2026xul}, ultraheavy cosmic rays
\citep{Patel:2026ynu} and the kilonovae of repeated flares in neutron-star---white-dwarf mergers \citep{Zhong:2026loz}.

In the shocked crust, electron and positron captures and the emission and absorption of neutrinos can change
$\Ye$. The time on which the captures alone change it falls steeply with temperature and the conditions under which an $\alpha$-rich freeze-out makes heavy nuclei
are very sensitive to $\Ye$. These processes have not yet been followed in simulations of
crust ejection. How much the neutrinos change the electron fraction of the ejecta, and with it their yields, can
only be found by evolving them together with the flow, and this is the main motivation of this work.

In this paper we study the crust-shock scenario with general-relativistic (GR) neutrino-radiation hydrodynamics simulations performed with the \athenak\ code \citep{stone2024athenakperformanceportableversionathena,Zhu:2024utz,Fields:2024pob}, using a nuclear equation of state (EOS) and energy-integrated two-moment (M1) neutrino transport, so that $\Ye$ of the ejecta evolves with the flow. We keep the 1D spherically symmetric geometry and the flare model of C24 and scan the flare pressure $\Pgf = B^2/8\pi$ of fields $B = 10^{15}$--$10^{16.5}\,$G, a range that brackets the fiducial field strength of
\citet{Patel:2025tse} and includes that of their high-mass model.
To isolate the effect of the neutrinos we repeat
every model without them. We do not vary GR and the EOS separately, so their effect appears only in the comparison with earlier work. The composition of the ejected crust is then computed with a nuclear reaction network along trajectories extracted from the simulations.

The paper is organized as follows. Section~\ref{sec:setup} describes the physical setup and Section~\ref{sec:methods} the numerical methods. Section~\ref{sec:results} presents the results. Section~\ref{sec:compare} compares them with the earlier calculations of C24 and \citet{Patel:2025tse}, and
Section~\ref{sec:discussion} discusses their implications and limitations.

\section{Physical setup}\label{sec:setup}

\subsection{Neutron star and flare layer}\label{sec:setup:star}

\begin{figure*}
	\centering
	\includegraphics[width=\textwidth]{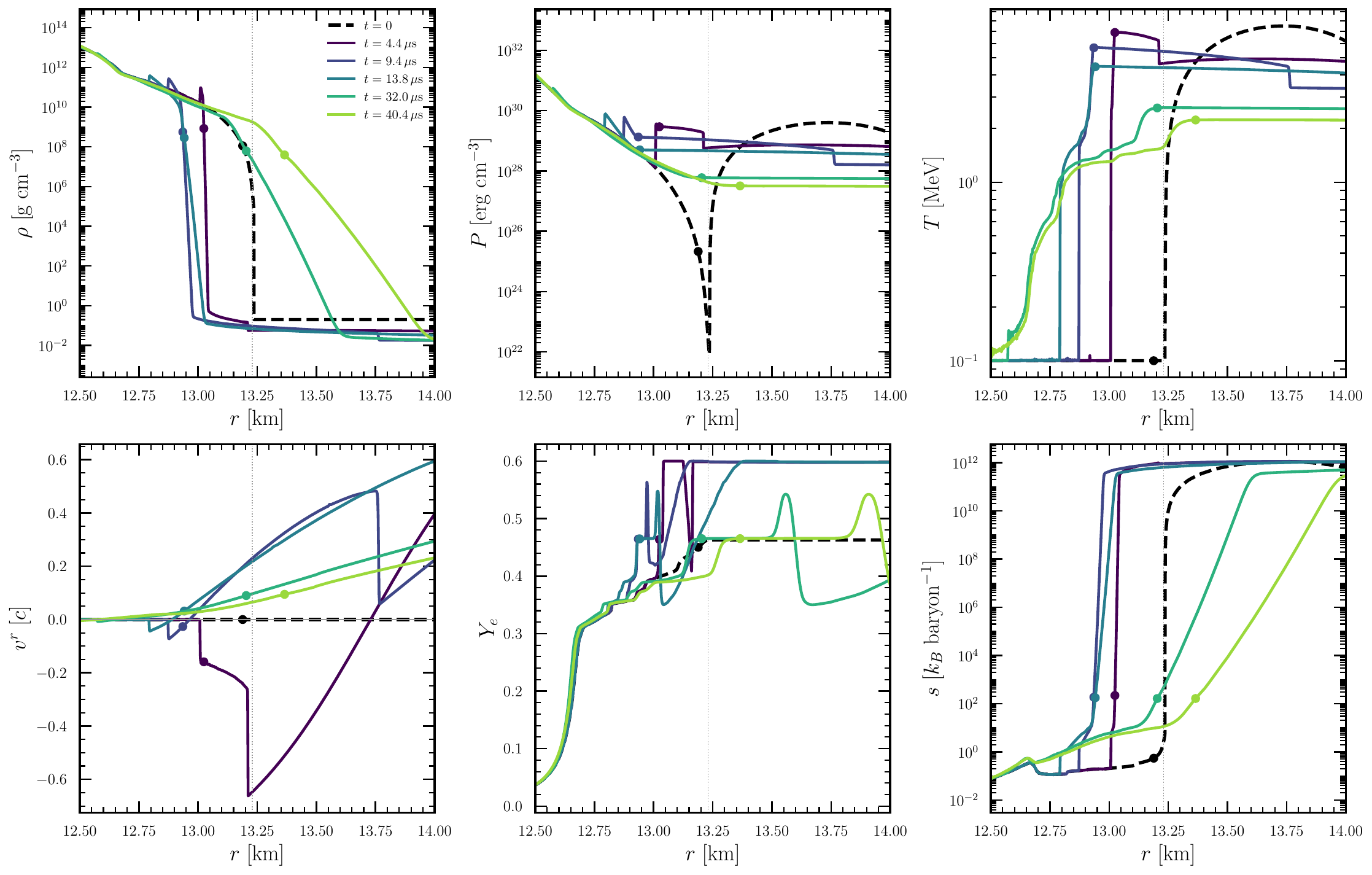}
	\caption{Radial profiles of model B15.5 near the stellar surface. \textit{Top}: rest-mass density, pressure and
		temperature. \textit{Bottom}: radial velocity, electron fraction and entropy per baryon. The black dashed curves show
		the initial state (crust and flare layer), the colored curves the five snapshots, three up to the deepest reach of
		the shock ($13.8\,\mu$s) and two while the crust flies out. Dots mark the radius outside which lies the gas
		that is eventually ejected, located with the mass coordinate of Equation~(\ref{eq:mass}). The thin vertical
		line marks the initial surface, $\Rns = 13.24\,$km.  Outside the crust, $\Ye$ of the dilute flare-layer gas rises to $0.6$, the upper limit of the EOS table. This gas is not part of the ejecta.}
	\label{fig:fiducial}
\end{figure*}

We construct a spherically symmetric NS at rest by solving the Tolman--Oppenheimer--Volkoff (TOV) equations
\citep{Tolman:1939jz,Oppenheimer:1939ne} for the $\beta$-equilibrated slice of the EOS (Section~\ref{sec:methods:eos}) at its lowest temperature, $T_{\rm min} = 0.1\,\MeV$. The central density $\rho_c = 5.86\times10^{14}\,\gcc$ gives a gravitational mass $\Mns = 1.40\,M_\odot$ and an areal
radius $\Rns = 13.24\,$km, where the surface is placed at $\rho = 10^{-10}\rho_c$. The crust is thus in GR hydrostatic equilibrium with the EOS used in the evolution. Figure~\ref{fig:fiducial} shows the initial profiles (dashed curves).

We evolve the region $12.0\,{\rm km} \leq r \leq 60\,$km for $1\,$ms ($2\,$ms without neutrinos), and every model is run both with and without neutrino transport. The shock stops well above the inner boundary (Section~\ref{sec:results:fiducial}). The mass above the inner boundary, $0.03\,M_\odot$, exceeds the ejecta masses of interest
by more than five orders of magnitude, so the evolved region contains all the matter that the flare can affect on this time scale.

We do not model the trigger of the flare. Its energy is thought to come from the magnetic field, whose slow evolution inside the star stresses the crust and twists the magnetospheric field lines
\citep{Thompson:1995gw,Thompson:2001ie}. The stored energy is then released suddenly, through a failure of the
crust under magnetic stresses \citep{Perna:2011gt,Lander:2014csa} or through a large scale reconnection instability of the twisted magnetosphere \citep{Lyutikov:2005un,Parfrey:2013gza}. In the relativistic magnetohydrodynamics simulation of such an eruption by \citet{Chatterjee:2026tdp}, reconnection
launches a hot outflow that can power the initial spike, while hot plasma falls back, reflects off the stellar surface and settles into a fireball trapped in the closed magnetosphere, with enough thermal energy to power the pulsating tail \citep{Thompson:1995gw}. That outflow carries no baryons, and the baryonic
ejecta of the 2004 flare may require another mechanism \citep{Chatterjee:2026tdp}, for instance the eruption of a magnetic loop that carries crustal material out of a young, strongly magnetized star \citep{Bransgrove:2025tpa}, or the scenario of C24 adopted here, in which the pressure of the hot layer drives a shock into the crust. The flare can also launch waves into the magnetosphere that steepen into strong shocks far from the star \citep{Grehan:2026vof}, but we do not model the magnetosphere and consider only the hot layer in contact with the crust.

At $t = 0$ we place a layer of hot plasma on the NS surface, with a peak pressure set by the magnetic pressure,
\begin{equation}
  \Pgf \simeq \frac{B^2}{8\pi}=4\times10^{28}\,{\rm erg\,cm^{-3}}\left(\frac{B}{10^{15}\,{\rm G}}\right)^2 .
  \label{eq:Pgf}
\end{equation}
Here $B$ labels the pressure of the layer and need not be the dipole field of the star. Equation~(\ref{eq:Pgf}) assumes that the full magnetic pressure is converted into plasma pressure. In force-free simulations of magnetospheric eruptions 15--45 percent of the magnetic free energy is dissipated, depending on how the field is sheared, and the rest is expelled with the erupting field or retained~\citep{Parfrey:2013gza}. In the relativistic magnetohydrodynamic simulations of \citet{Chatterjee:2026tdp} the trapped fireball keeps most of its energy in the magnetic field, with a plasma pressure well below $B^2/8\pi$. For a given field Equation~(\ref{eq:Pgf}) gives the highest pressure the layer can have.
The layer extends from $\Rns$ to $\Rns + 1\,$km, the fiducial width of C24, and its pressure has a
raised-cosine profile,
\begin{equation}
  P(r) = \frac{\Pgf}{2}\left[1 + \cos\left(\pi\,\frac{|r - r_c|}{w}\right)\right], \qquad |r - r_c| < w ,
  \label{eq:fireball}
\end{equation}
with $r_c = \Rns + 0.5\,$km and $w = 0.5\,$km. The dissipation region is about one scale height of the
magnetospheric energy density thick, so its pressure should fall off smoothly. A smooth profile also avoids a
pressure jump of five to ten orders of magnitude at the stellar surface, whose early evolution would depend on the
grid spacing. We vary the geometry of the layer in Section~\ref{sec:results:layer}.

The layer is initially at rest, with density $\rho_{\rm fb} = 0.2\,\gcc$ and the electron fraction of the
stellar surface, $\Ye = 0.463$. With the $\Ye$ of the
surface there is no jump in composition at the interface with the crust, and numerical diffusion across it cannot
change $\Ye$ of the crust. Its temperature follows from the EOS (Table~\ref{tab:models}). The layer carries as little rest mass as possible ($3\times10^{-16}\,M_\odot$ in
total), with a density $100$ times that of the atmosphere, which keeps it well above the density floor of the code (Section~\ref{sec:setup:atm}) while it drives the shock into the crust. Its pressure is
dominated by radiation and $e^\pm$ pairs, with a nucleon contribution $\lesssim10^{-8}$. Its rest-mass energy,
$5\times10^{38}\,$erg, is less than $10^{-8}$ of its thermal energy $E_{\rm fb}$ (Equation~\ref{eq:Efb}), so the value of $\rho_{\rm fb}$ does not set the energy
of the flare.
 We define the energy of the layer as its thermal energy above the cold state,
\begin{equation}
  E_{\rm fb} = \int_{\rm layer} \rho\,\left[\epsilon(\rho, \Ye, T) - \epsilon(\rho, \Ye, T_{\rm min})\right]
  \sqrt{\gamma}\,d^3x ,
  \label{eq:Efb}
\end{equation}
where $\epsilon$ is the specific internal energy, $T$ the temperature, and $\sqrt{\gamma}\,d^3x$ the proper volume element, with $\gamma$ the determinant of the
spatial metric (Section~\ref{sec:methods:hydro}).

Because the layer covers the whole star, $E_{\rm fb}$ (Table~\ref{tab:models}) is the energy of a flare that
dissipates $\Pgf$ over the entire surface. Our models
exceed the 2004 spike, $(2$--$4)\times10^{46}\,$erg (Section~\ref{sec:intro}), by factors of 4 (B15.0) to $4\times10^{3}$ (B16.5). A flare that
covers a solid angle $\Delta\Omega$ holds only $(\Delta\Omega/4\pi)\,E_{\rm fb}$ at the same pressure, and the
region where the field opens and matter can escape may be small. We include flares of this energy because the
delayed MeV emission of the 2004 flare implies an ejecta mass of ${\sim}10^{-6}\,M_\odot$ \citep{Patel:2025frn}.
They also span the field strengths of C24 and \citet{Patel:2025tse}, which allows a direct comparison with their
results, and they show how the ejecta depend on $\Pgf$ over a wide range, even where a flare of that energy over
the whole surface is unlikely.

\begin{deluxetable}{ccccc}
\tablecaption{Models are named after $\log_{10}(B/{\rm G})$. All models share the star, grid and physics and
differ only in $\Pgf$ (Equation~\ref{eq:Pgf}). $T_{\rm peak}$ is the largest temperature of the layer at
$t = 0$, and $E_{\rm fb}$ its energy (Equation~\ref{eq:Efb}).\label{tab:models}}
\tablehead{\colhead{Model} & \colhead{$\log_{10} B$} & \colhead{$\Pgf$} & \colhead{$T_{\rm peak}$} &
\colhead{$E_{\rm fb}$} \\
\colhead{} & \colhead{[G]} & \colhead{[erg cm$^{-3}$]} & \colhead{[MeV]} & \colhead{[erg]}}
\startdata
B15.0 & 15.0 & $4.0\times10^{28}$ & 4.23 & $1.7\times10^{47}$ \\
B15.25 & 15.25 & $1.3\times10^{29}$ & 5.63 & $5.5\times10^{47}$ \\
B15.5 & 15.5 & $4.0\times10^{29}$ & 7.51 & $1.7\times10^{48}$ \\
B15.75 & 15.75 & $1.3\times10^{30}$ & 10.0 & $5.4\times10^{48}$ \\
B16.0 & 16.0 & $4.0\times10^{30}$ & 13.4 & $1.7\times10^{49}$ \\
B16.25 & 16.25 & $1.3\times10^{31}$ & 17.8 & $5.4\times10^{49}$ \\
B16.5 & 16.5 & $4.0\times10^{31}$ & 23.7 & $1.7\times10^{50}$ \\
\enddata

\end{deluxetable}

\subsection{Atmosphere}\label{sec:setup:atm}

The rest of the domain is filled with a uniform atmosphere at rest, with $\rho = 2\times10^{-3}\,\gcc$,
$T = T_{\rm min}$ and $\Ye = 0.463$. Its density is just above the lowest density of the extended EOS table,
$1.7\times10^{-3}\,\gcc$ (Section~\ref{sec:methods:eos} and Appendix~\ref{app:eos}).
This density and $T_{\rm min}$ are also the floors of the code. The atmosphere is not in hydrostatic equilibrium, and its rest mass in the domain,
$\approx1\times10^{-15}\,M_\odot$, is negligible compared with the ejecta. Its inertia is larger. At $T_{\rm min}$ photons and pairs give it an
enthalpy density $e + P \approx 2\times10^{22}\,{\rm erg\,cm^{-3}}$, with $e$ the energy density without rest mass,
so the ejecta sweep up an effective mass $(e + P)/c^2 \approx 1\times10^{-12}\,M_\odot$ within the extraction radius
$R_{\rm ext} = 30\,$km (Section~\ref{sec:methods:domain}). This is about
1 percent of the ejecta of B15.0 and less for the stronger flares, but it is comparable to the ejecta of weaker
flares. In test runs with $B = 10^{14.5}\,$G the ejecta are slowed down by the atmosphere, and with $10^{14}\,$G the
crust stalls at about $33\,$km. We therefore restrict the study to $B \geq 10^{15}\,$G. A control run without
the flare layer ejects only $5\times10^{-17}\,M_\odot$ of atmosphere through $R_{\rm ext}$ in $1.35\,$ms, six orders of
magnitude below B15.0, and the contact with the atmosphere warms the top $\sim150\,$m of the crust to at most
$0.46\,\MeV$, negligible compared with the shock heating.

\section{Numerical methods}\label{sec:methods}

\subsection{General-relativistic hydrodynamics}\label{sec:methods:hydro}

We solve the equations of GR hydrodynamics in the 3+1 ``Valencia'' formulation
\citep{Banyuls:1997zz,Anton:2005gi} with the dynamical-spacetime GRMHD module \citep{Fields:2024pob} of \athenak, a
performance-portable rewrite of Athena++ based on Kokkos \citep{stone2024athenakperformanceportableversionathena,9485033}. The evolved variables
are the conserved rest-mass density $D = \rho W$, the momentum density $S_i$ and the energy density
$\tau = E - D$, where $\rho$ is the rest-mass density in the fluid frame, $W$ the Lorentz factor and $E$ the energy
density measured by Eulerian observers.
The electron fraction is evolved as $D\Ye$, with the neutrino source terms of Section~\ref{sec:methods:nu}.
The magnetic field is zero.

The spacetime is fixed to that of the initial TOV star (Cowling approximation), with zero shift. The evolved region, from $12\,$km to the surface, holds less than 3 percent of the stellar mass, and the gas that the flare displaces, the ejecta and the lifted bound crust, is $\lesssim2\times10^{-5}$ of it. The line element is $ds^2 = -\alpha^2 dt^2 + \gamma_{ij}dx^i dx^j$, with lapse
$\alpha(r)$ and spatial metric $\gamma_{ij}$. The domain is a thin radial patch with
$\gamma_{rr} = (1 - 2m_g(r)/r)^{-1}$, where $m_g(r)$ is the gravitational mass inside areal radius $r$,
$\gamma_{yy} = \gamma_{zz} = r^2$ and a single periodic cell in each transverse direction, which reproduces
the geometric source terms of spherical symmetry. The fluid four-velocity is $u^\mu$, with $W = \alpha u^t$ and
radial three-velocity $v^r = u^r/W$.

We use fifth-order WENO-Z reconstruction \citep{2008JCoPh.227.3191B} of the density, $W v^i$ and $\Ye$ (the pressure is
reconstructed by the well-balanced scheme below), the HLLD Riemann solver
\citep{Mignone:2008ii,Fields:2026gay}, and third-order strong-stability-preserving
Runge--Kutta time integration~\citep{doi:10.1137/S003614450036757X} with a Courant number of 0.4. The crust spans many orders
of magnitude in density over a few hundred meters, and a standard finite-volume scheme does not preserve its
discrete hydrostatic balance. We employ a locally well-balanced scheme designed to preserve hydrostatic equilibria
(Fields \& Dittmann, in preparation) to reconstruct
the pressure. This scheme and the FOFC are described in Appendix~\ref{app:numerics}. A
first-order flux correction \citep[FOFC;][]{Lemaster:2008gh,Fields:2024pob} keeps the evolution robust. Cells that fall below the density floor are reset to the atmosphere at rest, and gas that cools below
$T_{\rm min}$ is kept at $T_{\rm min}$ \citep{Fields:2024pob}, which adds thermal energy to dilute, expanding gas
(Section~\ref{sec:methods:domain}). The primitive
variables are recovered with the PrimitiveSolver library \citep{Cook:2023bag,Fields:2024pob}, based on
\citet{Kastaun:2020uxr}.

\begin{figure*}
	\centering
	\includegraphics[width=\textwidth]{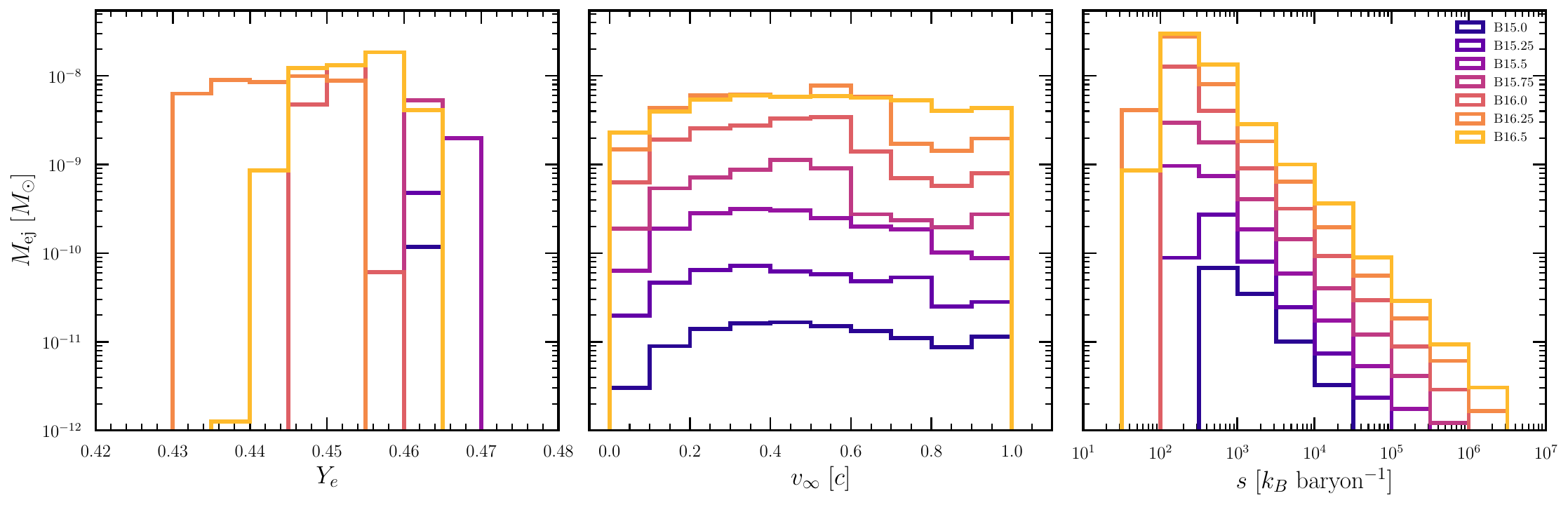}
	\caption{Mass distribution of the ejecta, integrated over the run, in electron fraction, asymptotic velocity
		$v_\infty$ and entropy per baryon at $R_{\rm ext}$, for all models, as the ejected mass in each bin. The mass-weighted
		averages and the median entropy are listed in Table~\ref{tab:results}.}
	\label{fig:hist}
\end{figure*}

\subsection{Equation of state}\label{sec:methods:eos}

We use the HS(DD2) EOS \citep{Hempel:2009mc,Typel:2009sy}, a statistical model of nuclei with measured masses \citep{Audi:2002rp} where available and masses from the finite-range droplet
model (FRDM, \citealt{Moller:1996uf}) otherwise, and of nucleons with the DD2 relativistic mean-field interaction, including electrons,
positrons and photons, in the CompOSE table \citep{CompOSECoreTeam:2022ddl} covering $T = 0.1$--$158\,\MeV$, baryon number
densities $n_b = 10^{-12}$--$10\,{\rm fm^{-3}}$ and $\Ye = 0.01$--$0.60$. Its lowest density, $1.7\times10^{3}\,\gcc$,
is far above that of the expanding flare layer and of the atmosphere, so we extended the table down to
$n_b = 10^{-18}\,{\rm fm^{-3}}$ ($1.7\times10^{-3}\,\gcc$) by continuing it as an ideal gas (adiabatic index $\Gamma = 5/3$) plus
a density-independent radiation (photon and pair) component, both matched to the table at its edge. The chemical
potentials are continued as well, since they enter the neutrino rates (Appendix~\ref{app:eos}).

The table describes matter in nuclear statistical equilibrium (NSE), with a distribution of nuclei, $\alpha$
particles and free nucleons. The dissociation of the crust nuclei by the shock, and the energy released when
nucleons recombine into $\alpha$ particles during the expansion, $\approx7\,\MeV$ per nucleon, are therefore part
of the evolution. Below
$T \approx 0.5\,\MeV$ the true composition freezes out, while the table keeps assuming NSE down to its lowest
temperature. The further energy that NSE releases by assembling heavier nuclei is at most
$\approx1.8\,\MeV$ per nucleon (from $\alpha$ particles to iron-group nuclei), about one percent of the
asymptotic kinetic energy of the ejecta, so the dynamics is not affected. The composition of the ejecta is
computed separately with a reaction network (Section~\ref{sec:methods:nuc}).

\subsection{Neutrino transport}\label{sec:methods:nu}

Neutrinos are evolved with the energy-integrated (gray) two-moment formalism \citep{Thorne:1981nvt,Foucart:2022bth}
with the Minerbo closure \citep{1978JQSRT..20..541M}, in the \athenak\ implementation ported from \texttt{THC\_M1}
\citep{Radice:2021jtw,Qiu:2026aop}. For each of four
species, $\nu_e$, $\bar\nu_e$ and the heavy-lepton $\nu_x$ and $\bar\nu_x$ (where $x$ combines the muon and tau (anti-)neutrinos), the scheme evolves the energy, momentum
and number densities. The radiation--matter source terms, set by the absorption and scattering opacities $\kappa_a$
and $\kappa_s$, act on the fluid energy, momentum and lepton number, and are integrated implicitly where the
neutrinos are trapped \citep{Radice:2021jtw}. Emissivities and opacities are
computed with \nurates\ \citep{Chiesa:2024lnu}, averaged over a neutrino spectrum reconstructed from the evolved
energy, number densities and the Eddington factors. They include charged-current absorption and emission on nucleons and (inverse) nucleon
decays \citep{Bruenn:1985en,Oertel:2020pcg}, with weak-magnetism and recoil corrections \citep{Horowitz:2001xf},
$e^+e^-$ pair processes \citep{Bruenn:1985en,Pons:1998st}, nucleon--nucleon bremsstrahlung
\citep{Hannestad:1997gc,Fischer:2016boc}, elastic scattering on nucleons \citep{Bruenn:1985en,Horowitz:2001xf} and
inelastic scattering on $e^\pm$ \citep{Bruenn:1985en,1993ApJ...410..740M}.
Where the optical depth over a time step exceeds unity, the thermal (excluding inelastic scattering) emissivities follow from the absorption opacities
through Kirchhoff's law, and elsewhere they are computed directly. Where the reconstructed rates are unphysical, i.e. the spectral temperature exceeds 2000 MeV, the rates fall back to assume an equilibrium spectrum at the predicted temperature and chemical potentials. The nucleon
rates neglect mean-field potentials and use the vacuum neutron--proton mass difference.

\subsection{Computational domain and diagnostics}
\label{sec:methods:domain}

The domain is covered by a base grid of $32\,$m cells with five nested refinement levels, all starting at
the inner boundary, with $16$, $8$, $4$, $2$ and $1\,$m cells out to $21.6$, $17.6$, $16.0$, $14.4$ and
$13.6\,$km, respectively. The whole crust and the inner part of the flare layer are thus resolved with $1\,$m cells (the rest of the
flare layer with $2\,$m cells). 
At the inner boundary the ghost cells are held at the initial hydrostatic state. In the innermost $200$m, $12.0$--$12.2$km, the conserved variables
are relaxed towards their initial values on a time scale of $0.05 \mu$s, which damps the waves that enter this layer
before they reach the boundary, and any wave the boundary reflects. The shock stops well above this layer. 
The outer boundary copies the state into the ghost cells and sets an inward velocity to zero, so that matter can leave the
domain but not enter it. The neutrino radiation field is zero at the inner boundary and copied at the outer boundary, without an inward flux.

We measure the ejecta on a sphere of areal radius $R_{\rm ext} = 30\,$km. Gas moving outwards
through it is counted as unbound if it satisfies the Bernoulli criterion
\begin{equation}
  -h\,u_t = h\,\alpha\,W \geq h_{\rm min},
  \label{eq:bern}
\end{equation}
where $u_t$ is the covariant time component of the four-velocity, $h$ the specific enthalpy (in units of $c^2$) and
$h_{\rm min} $ its minimum over the EOS table, the value to which cold, dilute
nuclear matter relaxes. The ejecta mass is the time integral of the mass flux of unbound gas through the sphere,
$\dot M = 4\pi r^2 \sqrt{\gamma_{rr}}\,\alpha \rho\,W v^r$, recorded every $0.25\,\mu$s. The asymptotic velocity
$v_\infty$ follows from the Bernoulli constant, $h\,\alpha\,W = h_{\rm min} W_\infty$, where
$W_\infty = (1 - v_\infty^2/c^2)^{-1/2}$ is the Lorentz factor the gas reaches far from the star, once its thermal
energy has been converted into motion. The electron fraction, $v_\infty$ and entropy of the ejecta are mass-weighted
over the same gas, from its state on the sphere in the output dumps, every $0.5\,\mu$s. The entropy of the ejecta has a long high-entropy tail (Section~\ref{sec:results:scan}), so we quote the
median entropy as well as the mean.

Equation~(\ref{eq:bern}) counts the thermal energy, mostly of photons and pairs, that dilute gas keeps only
because the code cannot cool it below $T_{\rm min}$. At $T_{\rm min}$ the table gives $h \simeq 1.5$ at $50\,\gcc$
and $1.16$ at $150\,\gcc$, while gas at rest at $30\,$km is unbound already for $h \gtrsim 1.08$, so dilute gas
held at $T_{\rm min}$ would count as unbound without moving outwards. We therefore measure the ejecta at
$R_{\rm ext} = 30\,$km, the largest radius at which no gas crossing the sphere has yet cooled to $T_{\rm min}$ in any
model.
If we remove from the gas that crosses $R_{\rm ext}$, cell by cell, the enthalpy it has at $T_{\rm min}$ at its density, so that only heating by shocks or neutrinos can unbind it, the ejecta mass decreases by 11 percent in B15.0 and by 3 percent or less in the other models.

\begin{figure*}
	\centering
	\includegraphics[width=0.8\textwidth]{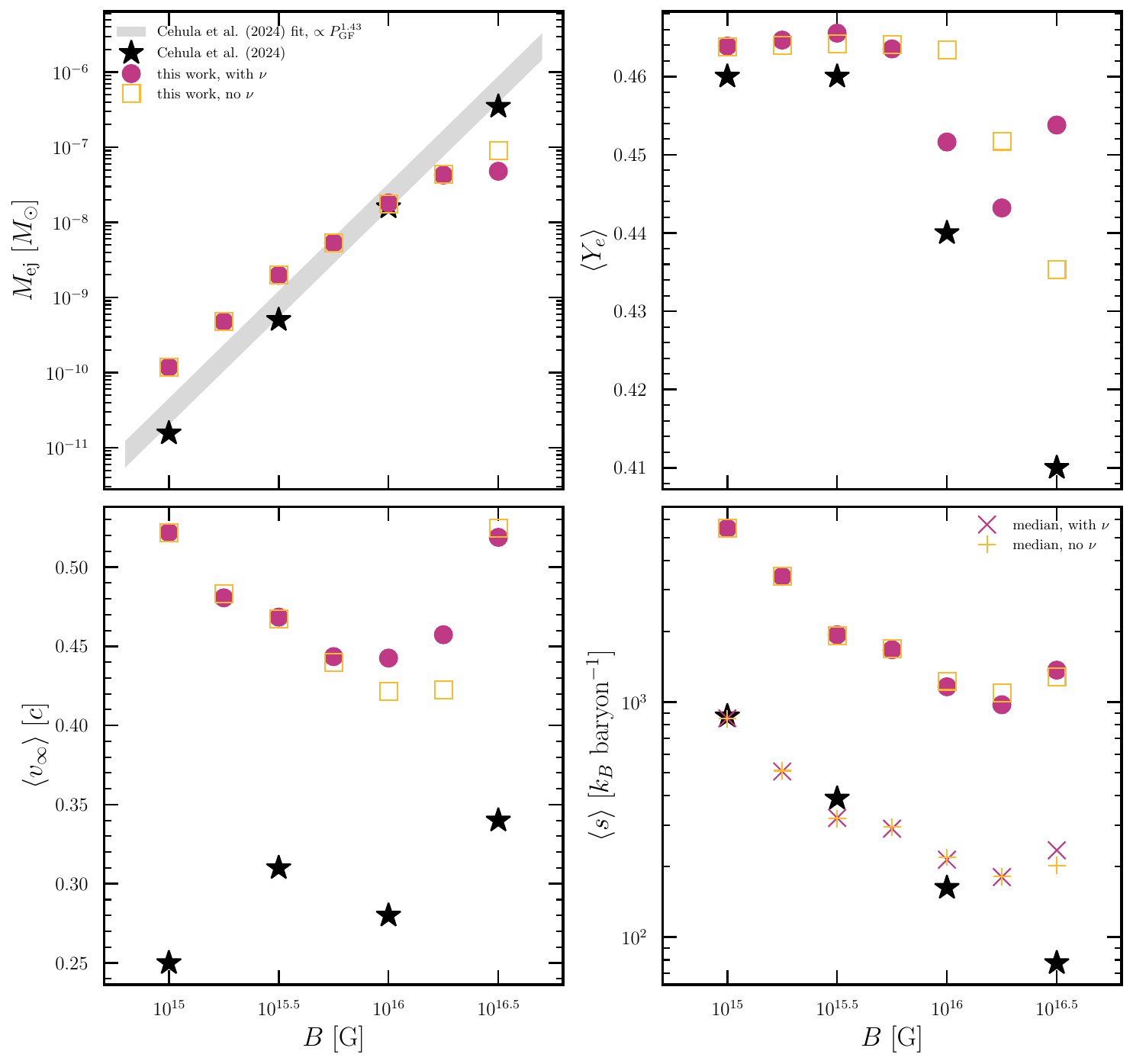}
	\caption{Ejecta of this work (filled circles: with neutrinos; open squares: without) against those of
		\citet[][C24; stars, their Table~1, 1 km layer]{Cehula:2023hdh}, as a function of $B$. In the upper left panel we show the ejecta mass, and the gray band is the
		fit of C24, $\Mej = (2$--$4.5)\times10^{-9}\,M_\odot\,(\Pgf/10^{30}\,{\rm erg\,cm^{-3}})^{1.43}$. Other panels:
		mass-weighted electron fraction, asymptotic velocity and entropy per baryon of the ejecta, with the mass-weighted
		median entropy (crosses), which is less sensitive to the fast, hot outermost layers
		(Section~\ref{sec:results:scan}). For C24, the mass-weighted means at the end of their runs. Our values are
		listed in Table~\ref{tab:results}.}
	\label{fig:c24}
\end{figure*}

\subsection{Nucleosynthesis}\label{sec:methods:nuc}

We compute the nucleosynthesis of the ejecta with the reaction network SkyNet \citep{Lippuner:2017tyn} along
Lagrangian trajectories extracted from the simulations. In spherical symmetry gas cannot overtake other gas, so the
rest mass outside a fluid element does not change as the element moves, and we use it to label the element.
Part of that mass leaves the grid through the outer boundary, so we count it as the mass between the element
and a sphere of radius $R_{\rm f}$ close to the boundary plus the net mass that has flowed out through this
sphere since $t = 0$,
\begin{equation}
  m(t, r) = \int_r^{R_{\rm f}} 4\pi r'^2 \sqrt{\gamma_{rr}}\,\rho W\,dr' + \int_0^t \dot M\,dt' ,
  \label{eq:mass}
\end{equation}
with $\dot M$ the mass flux of all the gas through $R_{\rm f}$, computed as in Section~\ref{sec:methods:domain}. At each output we find an element
at the radius where $m(t, r)$ takes its value. The gas that is eventually unbound is the outermost $\Mej$ of the
crust, that is, the gas with $m_{\rm s} < m < m_{\rm s} + \Mej$, where $m_{\rm s} = m(0, \Rns)$ is the value at the
crust surface, the rest mass of the flare layer and the atmosphere. We divide this gas into 147 shells, logarithmically spaced in
$m - m_{\rm s}$ in the outer half of $\Mej$ and of $0.005\,\Mej$ each in the inner half. In this way the
outer, high-entropy layers, which tracers of equal mass under-resolve \citep{Harris:2017gnl}, are resolved, as
is the deepest ejected gas, which makes most of the heavy nuclei. We follow the center of each shell through the output dumps, recording $\rho$, $T$, $\Ye$, the speed $v = \sqrt{\gamma_{rr}}\,v^r$ and the entropy per baryon $s$, averaged over the rest mass of the shell.

A trajectory starts when its shell cools through $9\,$GK for the last time, where we assume NSE at the local $(\rho, T, \Ye)$. Every ejected shell is heated above this
temperature by the shock. In B15.5 and B16.0 a second shock heats
deep shells (49 and 42 percent of the ejected mass) above $9\,$GK again after they first cooled through it,
and these shells start after this second heating. In B16.0 this shock comes from below, reaches the base of the
ejecta at $\approx220\,\mu$s and runs outwards through it, raising the entropy of the shells it crosses by 10--24
percent. The simulated part ends
when the shell reaches $50\,$km, before it approaches the outer boundary, or its temperature drops to $0.15\,\MeV$, above
the floor of the EOS table. There every shell is still at least 35 times denser than the density floor
(Section~\ref{sec:setup:atm}). Beyond this point, reached at radius $r_{\rm e} = 26$--$50\,$km, time $t_{\rm e}$,
temperature $T_{\rm e} = 0.15$--$0.51\,\MeV$, density $\rho_{\rm e} = 0.08$--$10^5\,\gcc$ and
velocity $v_{\rm e}$, the shell expands freely, $\rho = \rho_{\rm e}\,(r_{\rm e}/r)^3$, and we continue its radius in two
ways. In the first the shell coasts at $v_{\rm e}$, $r = r_{\rm e} + v_{\rm e}(t - t_{\rm e})$. In the second it
keeps the Bernoulli constant it has at $r_{\rm e}$ (Section~\ref{sec:methods:domain}), with
\begin{equation}
  h\,\alpha\,W = h_{\rm e}\,\alpha_{\rm e}\,W_{\rm e}, \qquad
  h - h_{\rm min} = \left(h_{\rm e} - h_{\rm min}\right)\frac{r_{\rm e}}{r},
  \label{eq:cont}
\end{equation}
and $\alpha = (1 - 2G\Mns/rc^2)^{1/2}$ of the exterior Schwarzschild metric. The second relation follows the thermal energy per baryon of a radiation-dominated gas in free expansion, which
falls as $T \propto 1/r$. The velocity then changes smoothly from $v_{\rm e}$ to the asymptotic velocity given by
$W_\infty = (h_{\rm e}/h_{\rm min})\,\alpha_{\rm e} W_{\rm e}$, as the remaining heat turns into motion and the shell
climbs out of the potential. A shell speeds up if $h_{\rm e}/h_{\rm min} > 1/\alpha_{\rm e}$ and slows down otherwise. A shell with $h_{\rm e}\alpha_{\rm e}W_{\rm e} < h_{\rm min}$ is
bound and is left out of the yields. These are the deepest shells of
B15.5, B16.0 and B16.5, with 5, 9 and 7 percent of the ejected mass, part of the gas that passes the Bernoulli
criterion at $R_{\rm ext}$ but fails it further out (Section~\ref{sec:results:scan}). Slower expansion keeps the
density, and hence the rate of seed formation and neutron captures, higher. We show the yields of the second
continuation and give the difference to the first in Section~\ref{sec:results:nuc}. Collisions between shells after the end of their simulated part are neglected. In both continuations at most two
shells per model move faster than the shell outside them, and they hold at most 3 percent of the ejected mass.

Along each trajectory SkyNet evolves the composition from NSE in two phases. During the simulated part, both the
temperature and the density follow the simulation, so that heating by later shocks, which the network cannot
compute, is included. After it, the density follows the continuation and the temperature follows from
the expansion and the nuclear energy release (self-heating), starting from the composition and temperature reached at
the end of the simulated part. We use SkyNet with the
Helmholtz EOS \citep{2000ApJS..126..501T}, strong and weak reaction rates from REACLIB \citep{2010ApJS..189..240C}, neutron-induced
fission with symmetric fragments \citep{Panov:2009yu}, spontaneous fission from the approximation of \citet{1947PhRv...72..914F} with the barriers of \citet{Mamdouh:2000ny}, the nuclear masses distributed with REACLIB, measured where available and from FRDM otherwise \citep{Moller:2015fba}, partition functions from \citet{Rauscher:2000fx}, and the screening distributed with SkyNet \citep{Lippuner:2017tyn}. We evolve each trajectory to $10^9\,$s. The yields of a model are the mass-weighted sums over its shells.

\section{Results}\label{sec:results}

\subsection{Dynamics of a fiducial flare}\label{sec:results:fiducial}
All models follow the same sequence. The flare layer drives a shock into the crust, the shock stops, the shocked crust decompresses outwards, and its outermost
layers are ejected. We describe this sequence for B15.5 and give the main numbers of every model in
Table~\ref{tab:dynamics}.

\begin{deluxetable}{lcccccc}
\tablecaption{Dynamics of the shock and origin of the ejecta. $T_{\rm cr}$: peak temperature of the shocked crust. $d_{\rm sh}$, $t_{\rm sh}$: depth below the
initial surface and time of the deepest reach of the shock. $P_{\rm cr}/\Pgf$: pressure of the initial crust there.
$d_{\rm ej}$, $\rho_{\rm ej}$: depth and initial density of the base of the ejecta, where the rest mass outside, in the
initial crust, equals $\Mej$. \label{tab:dynamics}}
\tablehead{\colhead{Model} & \colhead{$T_{\rm cr}$} & \colhead{$d_{\rm sh}$} & \colhead{$t_{\rm sh}$} &
\colhead{$P_{\rm cr}/\Pgf$} & \colhead{$d_{\rm ej}$} & \colhead{$\rho_{\rm ej}$} \\
\colhead{} & \colhead{[MeV]} & \colhead{[m]} & \colhead{[$\mu$s]} & \colhead{} & \colhead{[m]} & \colhead{[g cm$^{-3}$]}}
\startdata
B15.0 & 4.0 & 266 & 11.4 & 0.70 & 23 & $1.4\times10^{7}$ \\
B15.25 & 5.3 & 339 & 12.3 & 0.65 & 34 & $4.1\times10^{7}$ \\
B15.5 & 7.0 & 434 & 13.8 & 0.65 & 49 & $1.2\times10^{8}$ \\
B15.75 & 9.2 & 546 & 15.3 & 0.65 & 63 & $2.5\times10^{8}$ \\
B16.0 & 12.1 & 645 & 16.3 & 0.91 & 86 & $6.4\times10^{8}$ \\
B16.25 & 15.8 & 718 & 15.8 & 0.94 & 106 & $1.2\times10^{9}$ \\
B16.5 & 20.2 & 801 & 15.8 & 0.89 & 109 & $1.3\times10^{9}$ \\
\enddata

\end{deluxetable}

\begin{figure}
	\centering
	\includegraphics[width=\columnwidth]{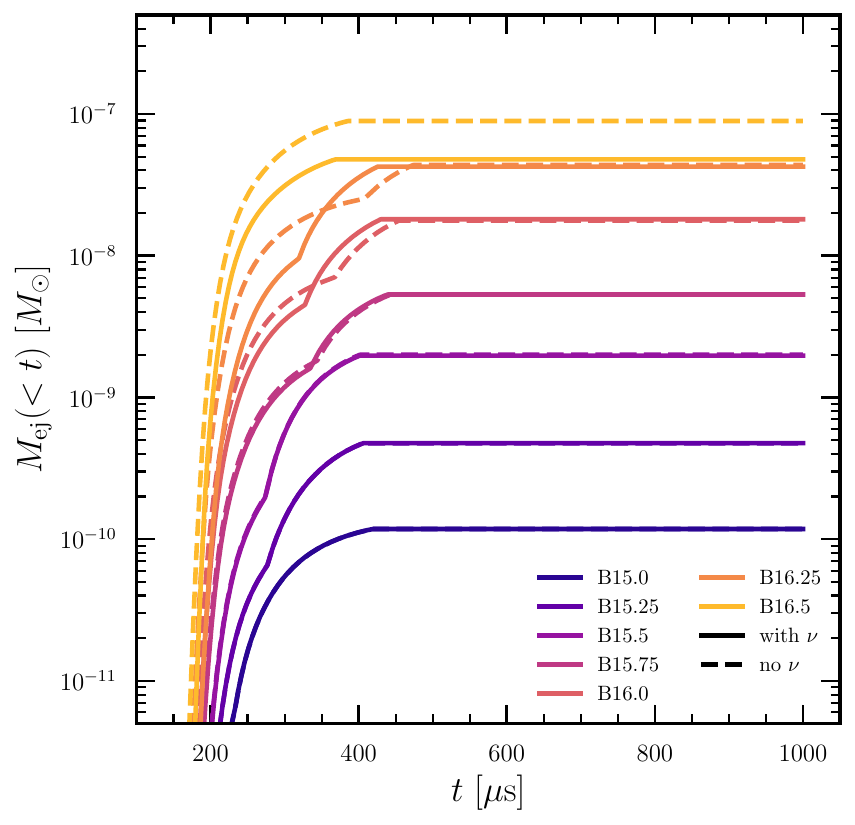}
	\caption{Unbound mass that has crossed $R_{\rm ext}$ by time $t$, with neutrinos (solid) and without (dashed),
			up to the common end time of each pair of runs.}
	\label{fig:cumulative}
\end{figure}

Figure~\ref{fig:fiducial} shows the first $40\,\mu$s of model B15.5. At $t = 0$ the flare layer sits on the cold
crust, at a peak pressure $\Pgf = 4\times10^{29}\,{\rm erg\,cm^{-3}}$ and temperature $7.5\,\MeV$. Its pressure
drives a shock into the crust, while the layer itself falls onto the star at up to $0.8\,c$. The shock heats
the outer crust to $T \approx 7\,\MeV$ and raises its entropy from $s < 1$ to ${\sim}10^{2}\,k_B$ per baryon
at the base of the ejecta. It slows down as it enters denser layers and reaches its deepest point after
$13.8\,\mu$s, at $12.80\,$km, $430\,$m below the surface, where the pressure
of the unperturbed crust is comparable to $\Pgf$.

As the flare layer expands, its pressure falls by an order of magnitude within $\approx15\,\mu$s, a rarefaction
enters the shocked crust and the crust decompresses outwards. By $\sim 30\,\mu$s the edge of the crust gas has
passed $13.5\,$km, and the outer crust moves at $\approx0.1\,c$, with the velocity increasing outwards. The
gas that is eventually ejected, comes from the outermost $\approx50\,$m of the crust (dots in Figure~\ref{fig:fiducial}). It starts with the electron fraction
of the outer crust, $\Ye \approx 0.45$--$0.46$, and ends with $\langle\Ye\rangle = 0.466$
(Section~\ref{sec:results:scan}), while the deeper, bound layers keep their lower $\Ye$. At the end of the run the
shocked crust is at $T \approx 0.7$--$1.4\,\MeV$. The ejecta cross $R_{\rm ext}$ between $\approx180$ and
$400\,\mu$s, the fastest first.

\begin{figure*}
	\centering
	\includegraphics[width=0.8\textwidth]{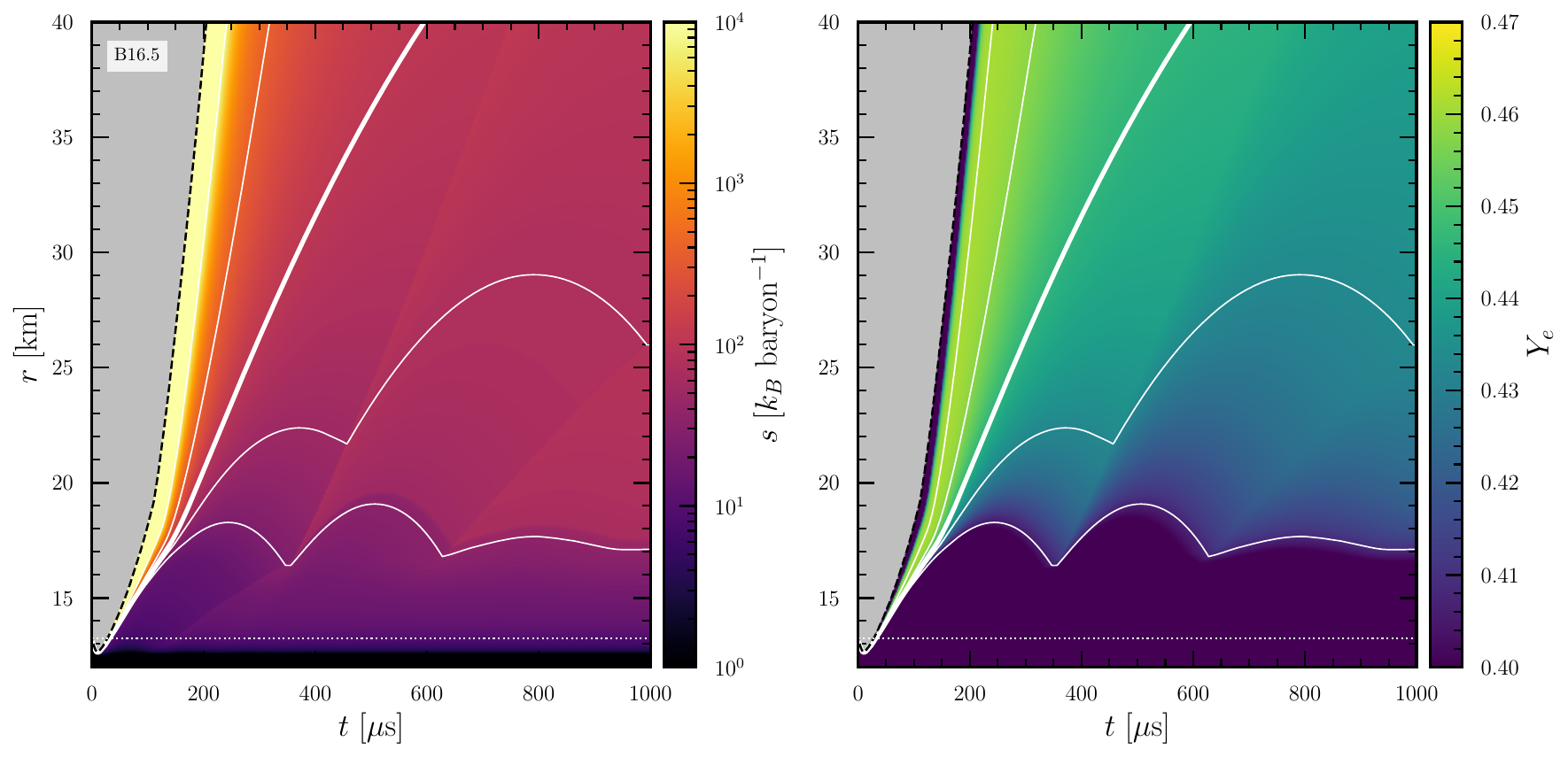}
	\caption{Space-time diagram of B16.5. Entropy per baryon (left) and electron fraction (right) of the crust gas
		against time and radius. The white lines are lines of constant rest mass outside them, $m = 0.01$ and
		$0.3\,\Mej$ in the ejecta and $2$ and $5\,\Mej$ in the bound crust, and the thick line is the base of the ejecta,
		$m = \Mej$. The dashed line is the contact between the flare layer and the crust, and the dotted line the
		initial surface of the star. The flare layer and the atmosphere are shown in grey. The thin dark band at the
		front of the ejecta in the right panel is a numerical effect. It
		holds less than $10^{-4}$ of $\Mej$ and makes no nuclei with $A \geq 90$.}
	\label{fig:spacetime}
\end{figure*}

Figure~\ref{fig:spacetime} shows the same sequence for B16.5 as a space-time diagram of the crust gas, with lines
of constant rest mass outside them that follow the gas. The flare layer pushes the crust down, and the shock heats it
until it stops, $800\,$m below the surface, after $15.8\,\mu$s (Table~\ref{tab:dynamics}). The crust then decompresses
outwards and its outer layers are ejected, the outermost first, while the bound gas below them falls back and oscillates. Each time it falls onto the denser layers below, a shock runs outwards through the bound crust, seen as the faint stripes in $s$ and $\Ye$ that leave the kinks of the lines in the bound crust. The entropy of the ejecta increases outwards,  and their electron fraction decreases with depth to $\approx0.43$--$0.44$ near the base (Section~\ref{sec:results:scan}).
In both models the flare thus ejects only the outermost $50$--$110$m of the crust, heated by a shock that lasts ${\sim}15\,\mu$s, while the deeper crust it compresses stays bound.

\subsection{Properties of the ejecta}\label{sec:results:scan}
\begin{deluxetable*}{lcccccccc}
\tablecaption{Ejecta and neutrino emission. $\Mej$: unbound mass (Equation~\ref{eq:bern}) through $R_{\rm ext}$, with neutrinos and, $\Mej^{{\rm no}\,\nu}$, without neutrinos, at $1\,$ms. $\langle\Ye\rangle$, $\langle v_\infty\rangle$, the median entropy $s_{1/2}$
and the kinetic energy at infinity $E_{{\rm kin},\infty} = \int (W_\infty - 1)\,c^2\,dM$, for all the gas and,
$E_{{\rm kin},\infty}^{<0.9c}$, for the gas with $v_\infty < 0.9c$. Averages and the median are mass-weighted. $E_\nu$: energy radiated in neutrinos of all species.\label{tab:results}}
\tablehead{\colhead{Model} & \colhead{$\Mej$} & \colhead{$\Mej^{{\rm no}\,\nu}$} &
\colhead{$\langle\Ye\rangle$} & \colhead{$\langle v_\infty\rangle$} &
\colhead{$s_{1/2}$} & \colhead{$E_{{\rm kin},\infty}$} & \colhead{$E_{{\rm kin},\infty}^{<0.9c}$} & \colhead{$E_\nu$} \\
\colhead{} & \colhead{[$M_\odot$]} & \colhead{[$M_\odot$]} & \colhead{} & \colhead{[$c$]} &
\colhead{[$k_B$ baryon$^{-1}$]} & \colhead{[erg]} & \colhead{[erg]} & \colhead{[erg]}}
\startdata
B15.0 & $1.2\times10^{-10}$ & $1.2\times10^{-10}$ & 0.464 & 0.52 & $8.5\times10^{2}$ & $2.5\times10^{44}$ & $4.4\times10^{43}$ & $5.1\times10^{42}$ \\
B15.25 & $4.8\times10^{-10}$ & $4.7\times10^{-10}$ & 0.465 & 0.48 & $5.1\times10^{2}$ & $7.7\times10^{44}$ & $1.7\times10^{44}$ & $7.4\times10^{43}$ \\
B15.5 & $2.0\times10^{-9}$ & $2.0\times10^{-9}$ & 0.466 & 0.47 & $3.2\times10^{2}$ & $2.5\times10^{45}$ & $6.6\times10^{44}$ & $1.1\times10^{45}$ \\
B15.75 & $5.3\times10^{-9}$ & $5.3\times10^{-9}$ & 0.464 & 0.44 & $2.9\times10^{2}$ & $7.3\times10^{45}$ & $1.4\times10^{45}$ & $1.6\times10^{46}$ \\
B16.0 & $1.8\times10^{-8}$ & $1.8\times10^{-8}$ & 0.452 & 0.44 & $2.1\times10^{2}$ & $2.2\times10^{46}$ & $4.8\times10^{45}$ & $2.1\times10^{47}$ \\
B16.25 & $4.3\times10^{-8}$ & $4.4\times10^{-8}$ & 0.443 & 0.46 & $1.8\times10^{2}$ & $5.6\times10^{46}$ & $1.2\times10^{46}$ & $2.2\times10^{48}$ \\
B16.5 & $4.8\times10^{-8}$ & $8.9\times10^{-8}$ & 0.454 & 0.52 & $2.3\times10^{2}$ & $1.2\times10^{47}$ & $1.9\times10^{46}$ & $2.1\times10^{49}$ \\
\enddata
\end{deluxetable*}
\begin{figure}
	\centering
	\includegraphics[width=\columnwidth]{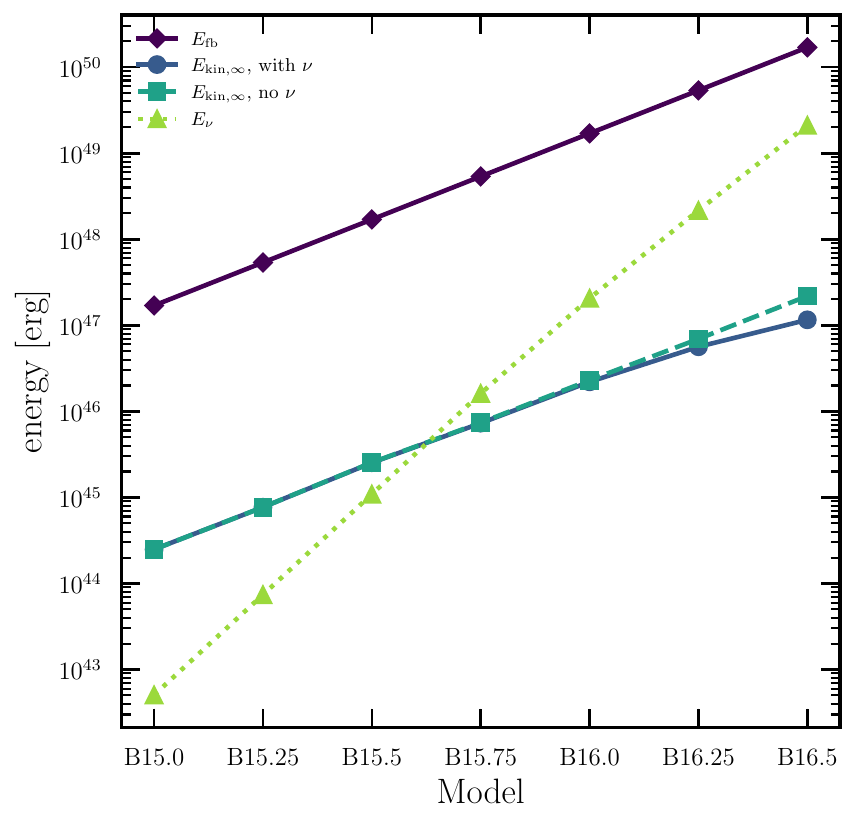}
	\caption{Energy of the flare layer
		(Equation~\ref{eq:Efb}), kinetic energy at infinity $E_{{\rm kin},\infty}$ of the ejecta
		with and without neutrino transport, and energy radiated in neutrinos as a function of the magnetic field $B$.}
	\label{fig:energy}
\end{figure}
\paragraph{Mass.}
Table~\ref{tab:results} summarizes the ejecta, and the upper left panel of Figure~\ref{fig:c24} shows their mass
against $B$. 
The flare unbinds between $1.2\times10^{-10}\,M_\odot$ (B15.0) and $4.8\times10^{-8}\,M_\odot$ (B16.5) of
crust. Between B15.0 and B16.0, the mass grows as $\Mej \propto \Pgf^{1.1}$, and it
flattens to $\Mej \propto \Pgf^{0.4}$ between B16.0 and B16.5, because neutrinos remove 46 percent of the mass of
B16.5. Without neutrinos B16.5 ejects $9\times10^{-8}\,M_\odot$ and the slope is $0.7$, and the masses of the other models
change by less than 5 percent. The ejection is over by $500\,\mu$s. With and without neutrinos no unbound gas crosses $R_{\rm ext}$ after that.

The ejecta mass also tells how deep the ejection reaches. In a thin crust in hydrostatic equilibrium the rest mass
above a layer is proportional to its pressure, $M_{\rm cr} \approx 4\pi\Rns^4 P_{\rm cr}/(G\Mns)$, so $\Mej$ fixes the
pressure $P_{\rm ej}$ of the initial crust at the base of the ejecta (Table~\ref{tab:dynamics}). In all models it is a few $10^{-5}\,\Pgf$.

These masses are measured at a finite radius with the Bernoulli criterion, which assumes a steady flow and tends
to overestimate the unbound mass \citep{Foucart:2021ikp}, so part of the gas that passes it at $R_{\rm ext}$ fails it
further out. Runs of B15.5, B16.0 and B16.5 without neutrinos with the outer boundary at $120\,$km instead of
$60\,$km give the same mass at $R_{\rm ext}$ to within 0.5 percent, so the outer boundary does not affect it. In these
runs the mass that passes the criterion still falls with radius, by about 10 percent from $30$ to $55\,$km, but only
by a further 3--6 percent out to $110\,$km. Our ejecta masses are therefore slight overestimates. Extrapolating the masses at $30$ and $55\,$km as $\Mej(r) = \Mej^\infty + a/r$ gives asymptotic masses 6--34 percent lower, an upper bound on the decrease, since at $110\,$km the runs with the larger domain lie above this law. The other properties of the ejecta hardly depend on the radius. Between $30$ and $55\,$km the mean asymptotic velocity changes by less than 5 percent, the median entropy by 9--11 percent for $B \geq 10^{15.5}\,$G, and the electron fraction not at all.

\paragraph{Distributions.}
Figure~\ref{fig:hist} shows the distributions of the ejecta. The gas keeps accelerating as its
remaining enthalpy is converted into kinetic energy, and its mass-weighted asymptotic velocities are
$0.44$--$0.52\,c$ (Table~\ref{tab:results}). The entropy spans several orders of magnitude. Most of the mass sits
at the low end, and the high-entropy tail comes from the outermost, fastest layers. Its leading edge is the
outermost skin of the crust, which has expanded below $1\,\gcc$ when it crosses $R_{\rm ext}$. It holds less than a few $10^{-4}$ of the ejecta for all models,
with $v_\infty \to c$ and entropies of $10^{6}$--$10^{9}\,k_B$ per baryon. Its density joins smoothly that of the
denser crust behind it, and the atmosphere swept up by the forward shock lies in a thin shell ahead of it. In every model 90 percent of the ejected mass
has $v_\infty = 0.10$--$0.96\,c$ and $s = 10^{2}$--$8\times10^{3}\,k_B$ per baryon, and its $\Ye$ lies within
$0.463$--$0.466$ up to B15.75, $0.448$--$0.455$ in B16.0, $0.433$--$0.453$ in B16.25 and $0.446$--$0.460$ in B16.5. The mass-weighted mean entropy, $\langle s\rangle = 1.0\times10^{3}$--$5.5\times10^{3}\,k_B$ per
baryon, is dominated by this tail and is $5$--$7$ times the mass-weighted median $s_{1/2}$ (Table~\ref{tab:results}).
Most of the ejecta cross $R_{\rm ext}$ in a single pulse (Figure~\ref{fig:cumulative}), with 10 and 90 percent of
the mass through the sphere between $220$ and $460\,\mu$s in every run. From B15.75 to B16.25 the
pulse has two maxima, 60--80$\,\mu$s apart. After the ejection the lifted crust
keeps rising and falling back with a period of $\approx100\,\mu$s (Section~\ref{sec:results:nu}), but this gas stays
bound (Figure~\ref{fig:spacetime}). In B15.5 and B16.0 the deepest ejecta are almost at rest ($\lesssim0.02\,c$) $150$--$200\,\mu$s after
the flare, when the lifted crust below them falls back onto the denser layers. The collision drives a second shock
outwards through the base of the ejecta, which accelerates it to $\approx0.3\,c$ and raises its entropy by up to 24
percent. In B16.5 the deepest ejecta still move out at $\approx0.2\,c$ at that time, and no such shock crosses them.
The distributions have similar shapes for all $B$.

\begin{figure}
	\centering
	\includegraphics[width=\columnwidth]{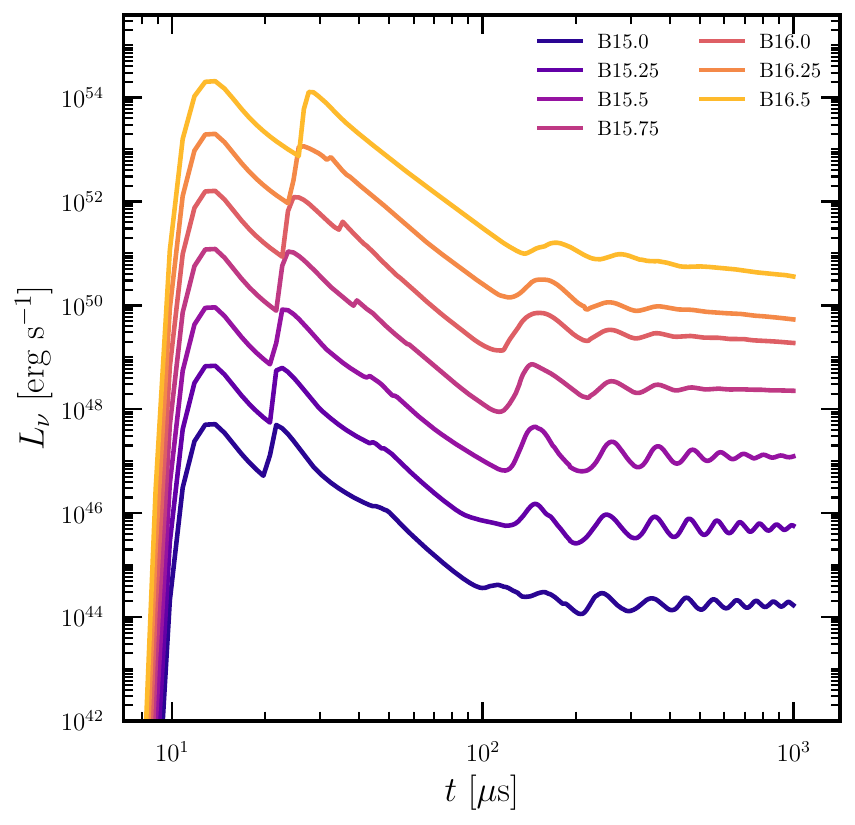}
	\caption{Total neutrino luminosity (all species) of all models.}
	\label{fig:nu}
\end{figure}

\paragraph{Energy.}
Figure~\ref{fig:energy} shows where the energy of the flare layer goes. Only a small and nearly constant fraction
ends up in the ejecta, $E_{{\rm kin},\infty} \approx (1.3$--$1.5)\times10^{-3}\,E_{\rm fb}$ from B15.0 to B16.0 and
$1.0\times10^{-3}$ in B16.25 and $6.8\times10^{-4}$$\,E_{\rm fb}$ in B16.5 (Table~\ref{tab:results}).

The energy per unit mass, $(W_\infty - 1)\,c^2$, grows without limit as $v_\infty \to c$, so most of the energy is
carried by a small fraction of the mass. The $88$--$95$ percent of the mass with $W_\infty < 2$ ($v_\infty < 0.87\,c$)
carries $14$--$24$ percent of $E_{{\rm kin},\infty}$. The fastest
tail is crust with the same $\Ye$ as the rest of the ejecta. It comes from the outermost, least dense layers, which
receive the highest energy per baryon, $\sim\Pgf/\rho \gg c^2$, expand as a baryon-loaded fireball and are the first
to cross $R_{\rm ext}$. 

The share of $E_{\rm fb}$ carried by neutrinos rises steeply with $B$, from $3\times10^{-5}$ in B15.0 to $0.12$ in B16.5 (Table~\ref{tab:results} and Figure~\ref{fig:energy}), so that from B15.75 on they carry more energy than the ejecta. This steep rise follows from the
temperature dependence of neutrino cooling. The thermal neutrino emission of a hot plasma rises steeply with
temperature, as $T^9$ for pair annihilation, while its thermal energy rises as $T^4$, so the cooling time falls
roughly as $T^{-5}$ and becomes comparable to the expansion time of the layer only in the strongest flares, whose
layers reach $13$--$24\,\MeV$ (Table~\ref{tab:models}). The neutrinos leave the layer at once because it is
transparent to them. The photons and pairs are trapped in the optically thick gas and move with it, and their energy can escape only later, as the observed $\gamma$-rays, once the fireball becomes transparent.
Up to B16.0 the neutrinos change the mass and the kinetic energy of the ejecta by less than 5 percent, and in B16.5 they halve it. In B15.5 and B16.0 the flare layer and the crust lose comparable amounts of energy to neutrinos. In B16.5 the net
loss comes entirely from the layer, which radiates $0.11\,E_{\rm fb}$, most of it within the first $10\,\mu$s, while it still drives the shock into the crust. The layer therefore pushes the crust with less energy and unbinds less of it.

Most of $E_{\rm fb}$ leaves with the flare-layer gas itself, for which the crust acts almost as a wall. In the runs without neutrinos 90--95 percent of the energy of the layer, measured as energy at infinity, streams outwards in every model, and the crust keeps the rest as heat and as the energy of the lifted, bound crust. 
With neutrinos the share is the same up to B16.0 and falls to $86$ and $78$
percent in B16.25 and B16.5, and the neutrinos carry most of the difference. This energy would leave as the fireball, ahead of the crust ejecta, and we did not measure
its effect on them.

\paragraph{Electron fraction.}

The ejected layers start with $\Ye = 0.43$--$0.46$ (Figure~\ref{fig:fiducial}). Without neutrinos the ejecta
keep about this value, $\langle\Ye\rangle = 0.463$--$0.464$ up to B16.0, slightly raised by numerical mixing at the
shock, and $0.452$ and $0.435$ for B16.25 and B16.5, whose ejecta come from deeper layers. Electron and positron
captures change $\Ye$ on a time scale $t_{\Ye} \simeq 2.3\,{\rm s}\,[\Ye/(1 - 2\Ye)]\,T_{\rm MeV}^{-5}$ (C24), which
falls from milliseconds at $4\,\MeV$ to microseconds at $20\,\MeV$. The shocked layers cool within ${\sim}10\,\mu$s, so
the captures matter only in the strongest flares, whose shocked crust reaches $12$--$20\,\MeV$. Up to B15.75
neutrinos change $\langle\Ye\rangle$ by at most $0.002$, and in B16.0 and B16.25 they lower it to $0.452$ and $0.443$.

In B16.5 the evolution is not monotonic. The captures first lower $\Ye$, but near the temperature peak the shocked
crust becomes so bright in neutrinos that the ejected layers reabsorb up to a third of the $\nu_e$ and $\bar\nu_e$
they emit. The imbalance between emission and absorption is at most one percent of either rate, yet the rates are so
high that it raises $\Ye$ by a few hundredths within a few microseconds. As the crust expands and cools the
absorption weakens and the captures lower $\Ye$ again. The ejecta end with $\langle\Ye\rangle = 0.454$, above the
value without neutrinos, $0.435$.

Whether the captures raise or lower $\Ye$ is decided by weak magnetism. Where the neutrinos escape, $\Ye$ relaxes
towards $Y_{e,{\rm eq}} = \lambda_{e^+n}/(\lambda_{e^+n} + \lambda_{e^-p})$, with $\lambda_{e^-p}$ and $\lambda_{e^+n}$
the capture rates per proton and per neutron. In the shocked crust thermal pairs outnumber the net electrons by
${\sim}10^2$, so electrons and positrons have nearly equal spectra, and since the neutron is heavier than the proton,
positron capture is faster. Alone, this would give $Y_{e,{\rm eq}} \approx 0.51$--$0.53$ and proton-rich ejecta. Weak
magnetism, the interference of the nucleon's anomalous magnetic moment with the axial current, has opposite signs for
neutrinos and antineutrinos. Together with nucleon recoil it lowers the rates of reactions that involve a $\bar\nu_e$
by $\approx(1 - 7E/m_N)$ and raises those that involve a $\nu_e$ by $\approx(1 + E/m_N)$, with $E$ the neutrino energy
and $m_N$ the nucleon mass \citep{Horowitz:1999fe,Horowitz:2001xf}. Positron capture emits a $\bar\nu_e$, and for
thermal leptons at $20\,\MeV$ its rate is halved, while electron capture changes by only 2--3 percent. $Y_{e,{\rm eq}}$
then drops to $\approx0.35$ at $20\,\MeV$ and $0.42$--$0.43$ at $10\,\MeV$, below the $\Ye$ of the crust, and the
captures lower $\Ye$ until they freeze out at $T \approx 4$--$6\,\MeV$. In supernova neutrino-driven winds the same
correction raises $\Ye$ \citep{Horowitz:1999fe,Horowitz:2001xf}, because there $\Ye$ is set by the absorption of
neutrinos, and the $\bar\nu_e$ reaction it weakens, $\bar\nu_e + p \to n + e^+$, turns protons into neutrons, whereas
in the shocked crust it weakens $e^+ + n \to p + \bar\nu_e$, which turns neutrons into protons. Runs of B16.0 and
B16.5 without weak magnetism and recoil end with $\langle\Ye\rangle = 0.514$ and $0.507$ instead of $0.452$ and
$0.454$, with ejecta masses within 8 percent of those of the main runs.  Without this correction the ejecta of the strongest flares would be proton-rich. With it, the ejecta keep an electron fraction close to that of the outer crust across the whole range of $B$, since the neutrinos change $\Ye$ by only $0.01$--$0.02$ from B16.0 upwards and by less below. They reduce the ejecta mass only in B16.5.

\subsection{Neutrino emission}\label{sec:results:nu}

\begin{figure}
	\centering
	\includegraphics[width=\columnwidth]{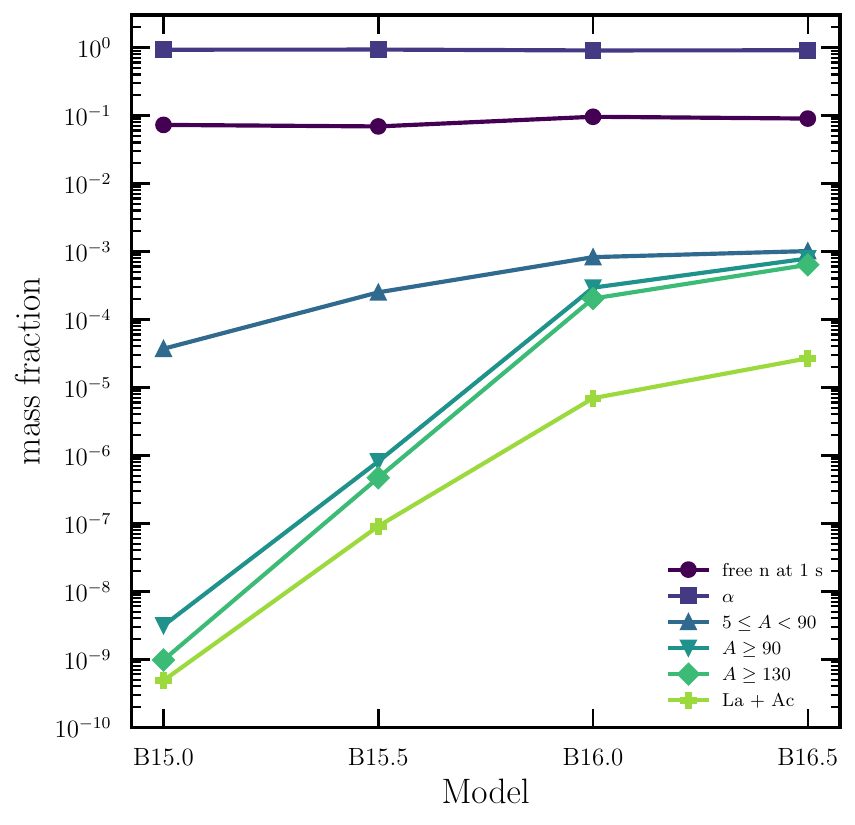}
	\caption{Mass fractions of free neutrons at $1\,$s, $\alpha$ particles, nuclei with $5 \leq A < 90$, $A \geq 90$,
		$A \geq 130$, and lanthanides plus actinides in the ejecta of B15.0, B15.5, B16.0 and B16.5, $10^{9}\,$s
		after the flare.}
	\label{fig:fractions}
\end{figure}

The neutrino luminosity $L_\nu$ (Figure~\ref{fig:nu}) peaks within the first ${\sim}20\,\mu$s, at
$L_\nu = 5\times10^{47}\,{\rm erg\,s^{-1}}$ in B15.0 and $2\times10^{54}\,{\rm erg\,s^{-1}}$ in B16.5, while
the flare layer is still hot and dense enough to emit and the shock heats the outer crust. 
The early emission has two peaks. The first peak comes from the flare layer and the second from the shocked crust. Both radiate at about the same time, but the crust lies deeper in the potential of the star, where light climbs out more slowly, so its neutrinos arrive $8$--$16\,\mu$s later, the more so for larger $B$, where the shock reaches deeper layers. By
$20\,\mu$s $3$--$40$ percent of the total $E_\nu$ has escaped and by $30\,\mu$s $60$--$80$ percent, most of it
emitted by the flare layer
(Section~\ref{sec:results:scan}). The luminosity then falls by factors of $500$--$6000$ by the end of the
runs, with peaks every $100$--$120\,\mu$s in every model, almost independent of $B$. From B15.75 on the peaks are smaller and die out after two to four cycles, by $\approx500\,\mu$s. The peaks come from the outer crust that is lifted but stays bound (Figure~\ref{fig:spacetime}). It falls back onto the denser layers below, which are already expanding again, and each collision drives a shock that heats the surface layers and gives a peak of $L_\nu$. \citet{Galeazzi:2013mia} found the same behavior in an oscillating neutron star, where gas falling back onto
re-expanding layers drives a shock at every oscillation and the luminosity of the outer layers follows the shock heating. The period is close to the dynamical time of the star,
$(\Rns^3/G\Mns)^{1/2} \approx 110\,\mu$s, and hardly depends on $B$. The luminosity then settles to a slowly decaying plateau as the heated crust cools. 

At the end of the runs it is still $L_\nu \approx 2\times10^{44}$, $1\times10^{47}$, $2\times10^{49}$ and
$4\times10^{50}\,{\rm erg\,s^{-1}}$ in B15.0, B15.5, B16.0 and B16.5. The crust thus keeps cooling by neutrinos after
$1\,$ms, and every further millisecond at this luminosity adds 2 (B16.5) to 11 percent (B15.5) of the energy emitted
during the run, so the values of $E_\nu$ in Table~\ref{tab:results} are lower limits on the total neutrino emission.  The neutrinos escape the flare layer directly, while the photons stay trapped in the fireball until it becomes
transparent, so their detection would probe the hot layer itself. Water Cherenkov detectors are most sensitive to the electron
antineutrinos, through inverse $\beta$ decay on free protons. B16.5 emits ${\sim}10^{53}$ of
them with a mean energy of ${\approx}50\,\MeV$, two thirds within the first $30\,\mu$s, and from $10\,$kpc they would arrive with a
fluence of ${\sim}10^{7}\,{\rm cm^{-2}}$. With the cross section of \citet{Strumia:2003zx} averaged over thermal
spectra, $\langle\sigma\rangle \approx 2\times10^{-40}\,{\rm cm^{2}}$, the expected number of events,
$N_{\rm ev} = N_p N_{\bar\nu_e}\langle\sigma\rangle/4\pi d^2$, is about two in Super-Kamiokande and $15$ in
Hyper-Kamiokande, whose fiducial masses of $22.5$ and $187\,$kt \citep{Hyper-Kamiokande:2018ofw} hold
${\sim}10^{33}$ and ${\sim}10^{34}$ free protons. Full conversion of the $\bar\nu_e$ into other flavors
would halve these numbers. The $\gamma$-ray spike, which rises in less than a millisecond and lasts ${\sim}0.2\,$s
\citep{Hurley:2005zs}, fixes the time of the flare. In the energy range of the positrons, $15.5$--$100\,\MeV$,
Super-Kamiokande expects ${\approx}2\times10^{-3}$ background events in $1000\,$s \citep{Super-Kamiokande:2018dbf},
and ${\approx}0.7$ when the range is extended down to $7\,\MeV$ \citep{Super-Kamiokande:2021dav}. Over the whole spike
the background is thus at most ${\sim}10^{-4}$ events, and even one or two events would be a clear detection. The number of events falls steeply with $B$, to ${\sim}10^{-2}$ in Super-Kamiokande for B16.0
($5\times10^{46}\,$erg in $\bar\nu_e$ with a mean energy of ${\approx}30\,\MeV$) and a few $10^{-5}$ for B15.5. Only the most extreme of our flares, far more energetic than the 2004 event, would thus give a detectable signal, and from giant flares like those observed we expect no neutrino events

\begin{figure*}
	\centering
	\includegraphics[width=0.95\textwidth]{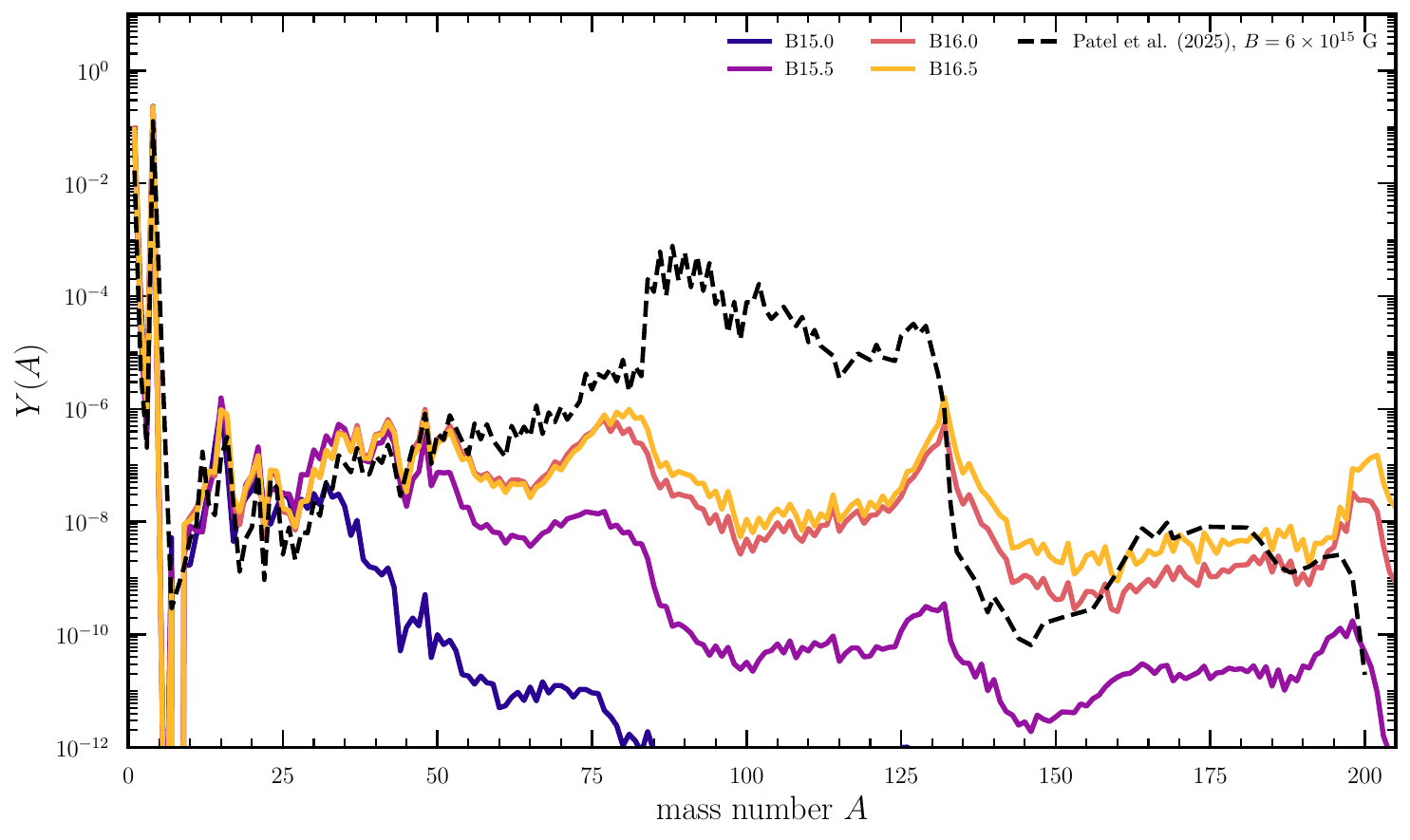}
	\caption{Mass-weighted abundances $Y(A)$ of the ejecta of B15.0, B15.5, B16.0 and B16.5, from SkyNet along the
		shell trajectories, $10^{9}\,$s after the flare, and, for comparison, the fiducial model of
		\citet[][their Figure~3, $B = 6\times10^{15}\,$G; dashed black]{Patel:2025tse}.}
	\label{fig:abund}
\end{figure*}

\subsection{Dependence on the flare layer}\label{sec:results:layer}

The shape, position and width of the flare layer are not constrained by observations, so we varied them at
fixed peak pressure and density, for B15.5 (shape) and for B15.5 and B16.0 (position and width). In all cases the
ejecta mass changes much less than the energy of the layer. It is set mainly by the peak pressure, which fixes how deep in the crust the ejection reaches (Section~\ref{sec:results:scan}).

A top-hat of the same peak pressure and width holds twice the energy of the raised cosine
($E_{\rm fb} = 3.4\times10^{48}\,$erg for B15.5) but ejects only 4 percent more mass. It pushes on the crust at the
full pressure $\Pgf$ from $t = 0$, and the extra energy goes mostly into the motion of the ejecta, which receive
$1.7$ times the kinetic energy at infinity and are slightly faster and hotter, with $\langle v_\infty\rangle = 0.50\,c$
against $0.47\,c$ and a median entropy of $360$ against $320\,k_B$ per baryon. Detaching the $1\,$km layer from the surface by $0.5$--$2\,$km of atmosphere raises its energy by up to 28 percent, as its volume grows, and the ejecta mass by at most 16 percent, while $v_\infty$, the median entropy and the time of the ejection hardly change (by at most 3 percent, 6 percent and $13\,\mu$s). Layers $0.5$, $2$ and $3\,$km wide, attached to the surface, hold about half, twice and three times the energy of the $1\,$km layer, but change the ejecta mass by less than a factor of two, with and without neutrinos, and without a monotonic trend. With neutrinos the $1\,$km layer ejects the most in both models, and the other widths eject $0.6$--$0.9$ times its mass. Our results thus depend on the flare mainly through its peak pressure, and the
unknown geometry of the flare layer changes the ejecta mass by less than a factor of two.

\subsection{Nucleosynthesis}\label{sec:results:nuc}

We show the nucleosynthesis of B15.0, B15.5, B16.0 and B16.5 (Figures~\ref{fig:abund} and~\ref{fig:fractions}).
Although the ejecta keep the moderately neutron-rich electron fraction of the crust, they make very few heavy nuclei. About 90 percent of the ejected mass ends as $\alpha$ particles, and the neutron excess, a mass fraction
$1 - 2\Ye \approx 0.07$--$0.10$, is left almost entirely as free neutrons. Nuclei with $A \geq 90$ make up from
$3\times10^{-9}$ of the ejecta in B15.0 to $8\times10^{-4}$ in B16.5, at most $3.5\times10^{-11}\,M_\odot$. Their
pattern extends to the third $r$-process peak, but lanthanides and actinides stay below $3\times10^{-5}$ of the mass. These fractions refer to the shells that escape in the network calculation, which hold more than 90 percent of the
ejected mass (Section~\ref{sec:methods:nuc}).

The heavy nuclei come from the deepest, densest shells. For $B \geq 10^{16}\,$G, the shells that start the network
above ${\approx}2\times10^{6}\,\gcc$ carry a fifth of the ejected mass and make 90 percent of the nuclei with
$A \geq 90$. These shells are also the least certain. The deepest 10 percent of $\Mej$ is barely unbound, and part of
it fails the Bernoulli criterion beyond $R_{\rm ext}$ (Section~\ref{sec:results:scan}). This slow, dense gas makes
$20$--$50$ percent of the nuclei with $A \geq 90$, and after the end of the simulated part the continuation expands it
somewhat more slowly than the simulation, which also favors heavy nuclei. The two continuations differ by up to a
factor of $1.4$ for $A \geq 90$, and by more for the heaviest nuclei. Our yields of heavy nuclei are therefore
probably upper limits, uncertain by up to a factor of two for $A \geq 90$ and by more for lanthanides and actinides,
while the bulk composition of $\alpha$ particles and free neutrons is not affected.

The free neutrons survive because there are almost no nuclei to capture them. Seed nuclei are assembled from
$\alpha$ particles mainly through $^4$He($\alpha n,\gamma$)$^9$Be($\alpha$,n)$^{12}$C, whose first step is a
three-body reaction, so that the seed production rate scales as $\rho^3$ while the gas cools from $T \approx 0.4$ to $0.2\,\MeV$ \citep{Hoffman:1996aj}. At fixed temperature the density falls as the entropy rises, and our
high-entropy ejecta form $\alpha$ particles at only ${\sim}10^{5}\,\gcc$ and expand through this phase within
${\sim}0.1\,$ms. Very few seeds form, and the density falls before the neutrons can be captured on them. Our ejecta thus lie at the extreme of an $\alpha$-rich freeze-out,
where the neutron excess stays in free neutrons because there are almost no nuclei to capture them. The crust
ejected by these flares therefore makes very few heavy nuclei, and its main radioactive product is the free
neutrons, $7$--$10$ percent of its mass (Section~\ref{sec:discussion}).

\section{Comparison with previous work}\label{sec:compare}

\subsection{Setup and physics}\label{sec:compare:setup}

Our calculations follow the flare model of C24, and our models bracket both the fiducial field of the
nucleosynthesis models of \citet{Patel:2025tse} (B15.5 and B16.0) and that of their strongest flare (B16.0 and B16.5,
their $B_{15} \equiv B/10^{15}\,{\rm G} = 13$).

To survey a wide range of flare energies and layer widths, C24 used special-relativistic hydrodynamics with a
$\Gamma = 4/3$ EOS, appropriate for the radiation-dominated shocked gas, on a crust in Newtonian hydrostatic
equilibrium, and estimated neutrino cooling and weak interactions in post-processing. Our setup differs from
theirs in the following points: 1) GR hydrodynamics on the fixed metric of the TOV star, where C24 used
a Newtonian point-mass potential. 2) A tabulated nuclear EOS with $\beta$-equilibrium composition, which also contains the dissociation of nuclei
by the shock and their recombination. C24 used a polytropic fit with
$\Gamma_* = 1.43$ for the initial crust. 3) M1 neutrino transport coupled to the fluid, so that $\Ye$ evolves. C24 estimated the weak interactions in
post-processing and assigned the ejecta the $\Ye$ of the crust at their mass coordinate. 4) A raised-cosine pressure profile for the flare layer, which at equal radius holds half the energy of the uniform layer
of C24 at the same peak pressure and width. With our larger radius and the curved-space volume, our $E_{\rm fb} \simeq 4\times10^{48}\,$erg at
$\Pgf = 10^{30}\,{\rm erg\,cm^{-3}}$ compares with $5\times10^{48}\,$erg in C24. 5) A smaller domain and a shorter evolution, $12$--$60\,$km for $1\,$ms with the ejecta measured at $R_{\rm ext}$,
against $10\,{\rm km} \le r \le 2\times10^4\,$km for $67\,$ms in C24. 6) An atmosphere at $T_{\rm min}$ (Section~\ref{sec:setup:atm}), where C24 attach a power law
$\rho \propto r^{-3}$ to the surface. 7) Cells of $1\,$m over the crust, against $2\,$m in C24.

\citet{Patel:2025tse} computed the nucleosynthesis with an analytic ejecta model motivated by the simulations of
C24. In our calculations the ejected mass and the thermodynamic history and $\Ye$ of each layer come from the
simulations.

\subsection{Ejecta}\label{sec:compare:ejecta}

At $B = 10^{16}\,$G our ejecta mass is close to that of C24 (B16.0 ejects $1.1$ times their mass). At B16.5 it is
$0.14$ times theirs ($0.26$ without neutrinos), and at B15.5 and B15.0 it is $3.9$ and $7.6$ times theirs, so that
it grows more slowly with $\Pgf$ than their $\Pgf^{1.43}$ (Figure~\ref{fig:c24}).

In C24 the ejecta reach crust densities with $\rho_{\rm cr}c^2 \approx (0.24$--$0.31)\,\Pgf$ at every $B$. Our weak
flares reach the same relative depth, $\rho_{\rm ej}c^2/\Pgf = 0.32$ and $0.27$ at B15.0 and B15.5, but at a given
density the DD2 outer crust, supported by relativistic degenerate electrons, has $3$--$4$ times the pressure of their
polytropic crust and as much more mass above it. This accounts for factors of $5.3$ and $4.2$, against the measured
$7.6$ and $3.9$. In strong flares the base of our ejecta is shallower, $\rho_{\rm ej}c^2/\Pgf = 0.14$ (B16.0) and
$0.03$ (B16.5), so that $\Mej$ grows nearly as $\Pgf$. The nuclear energy in our EOS, from the dissociation of nuclei
by the shock and their recombination during the expansion, changes the ejecta mass by at most a few percent, since
only 4--6 percent of the ejected mass has $(W_\infty - 1)\,c^2$ below $7\,\MeV$ per nucleon.

The electron fraction agrees with that of C24. Both keep the value of the crust up to B15.5 and decrease by a few
hundredths at higher $B$, in C24 because deeper, more neutron-rich layers are ejected and in our B16.0 through the
weak interactions (Section~\ref{sec:results:scan}).

Our ejecta are faster, $\langle v_\infty\rangle = 0.44$--$0.52\,c$ against $0.25$--$0.34\,c$. C24 showed that their
velocities depend on the numerical scheme. For their B15.0, parabolic reconstruction gives $0.47\,c$ and 5--6 times
the mass (their Table~2), close to our $0.52\,c$ and $1.2\times10^{-10}\,M_\odot$. Our mean entropies are $5$ (B15.5)
to $18$ (B16.5) times theirs because the dilute outermost layers, at $10^3$--$10^9\,k_B$ per baryon, dominate the mean
(Section~\ref{sec:results:scan}). Our median entropies are within 20 percent of their means up to B15.5, and $210$ and
$230\,k_B$ per baryon at B16.0 and B16.5 against their $160$ and $80$.

Neutrino losses reduce the ejecta mass appreciably only in B16.5, for $\Pgf \gtrsim 10^{31.5}\,{\rm
erg\,cm^{-3}}$. C24 found $\sim 2$ times more ejecta when the layer is widened from $1$ to $2\,$km, while between
$0.5$ and $3\,$km our ejecta change by less than a factor of two, without a monotonic trend (Section~\ref{sec:results:layer}). From B15.0 to B16.0 our ejecta masses thus lie within a factor of eight of those of C24, mostly because our crust holds more mass at a given density, and they fall below theirs at B16.5, where the ejection reaches a shallower depth and the neutrinos halve the ejecta.

\subsection{Nucleosynthesis}\label{sec:compare:nuc}

Our yields of heavy nuclei are three to four orders of magnitude below those of \citet{Patel:2025tse}. In their
fiducial model ($B = 6\times10^{15}\,$G) nuclei with $A \geq 90$ make up ${\sim}0.3$ of the ejected mass (their
Figure~3), mostly just above the first peak. In our B16.0 and B15.5, which bracket this field, they make up
$3\times10^{-4}$ and $8\times10^{-7}$. The strongest flare of the same work ($B = 1.3\times10^{16}\,$G) ejects
${\approx}6\times10^{-7}\,M_\odot$ of $r$-process nuclei, light ones included, almost twenty thousand times the
$3.5\times10^{-11}\,M_\odot$ with $A \geq 90$ of our B16.5.

The difference comes from the analytic ejecta profile that \citet{Patel:2025tse} adopt. At $B = 6\times10^{15}\,$G it
unbinds $1.1\times10^{-7}\,M_\odot$, $15$--$30$ times the mass C24 obtained in their simulations (their fit in
Figure~\ref{fig:c24}) and $6$--$60$ times that of our B16.0 and B15.5, so that their ejecta reach a mass depth ($m$ in
Equation~\ref{eq:mass}) of $10^{-7}\,M_\odot$, against $1.8\times10^{-8}$ (B16.0) and $2.0\times10^{-9}\,M_\odot$
(B15.5) in our models. Their escape criterion compares the enthalpy of each layer just behind the shock with
its binding energy, with the full flare pressure at every depth and each layer evolving on its own, and it reaches
crust with $\rho_{\rm cr}c^2 \approx 3\,\Pgf$. In the simulations the pressure of the flare layer drops within
${\sim}10\,\mu$s, the shock weakens as it moves into denser crust, and the deep layers give much of their energy to
the gas above them, which is why the fastest tenth of our ejecta carries most of the kinetic energy. In our runs the
crust that passes the Bernoulli test ${\sim}10\,\mu$s after the flare is three to twelve times the final ejecta mass,
and its deeper part becomes bound again within $10$--$150\,\mu$s. The ejecta finally come from crust with
$\rho_{\rm cr}c^2 \approx (0.03$--$0.3)\,\Pgf$ in our models and $\approx(0.2$--$0.3)\,\Pgf$ in those of C24.
In their profile the post-shock entropy falls with depth, down to $s \approx 20\,k_B$ per
baryon, and 82--98 percent of their ejected mass comes from this deeper, denser and colder crust, which makes most of
their heavy nuclei (their Figure~2). In our simulations the flare does not unbind these layers. The gas it ejects has
$s \gtrsim 100\,k_B$ per baryon down to the base of the ejecta, expands fast and freezes out rich in $\alpha$
particles with few seed nuclei (Section~\ref{sec:results:nuc}). The lower electron fraction of their ejecta,
$\Ye = 0.44$ against $0.44$--$0.47$ in ours, also favors heavier nuclei but plays a minor role. The large $r$-process yields of the analytic model therefore rest on the ejection of deep crust, which the flares in our simulations do not unbind.

\section{Discussion and conclusions}\label{sec:discussion}

We have followed the crust ejection of magnetar giant flares with general-relativistic hydrodynamics, a nuclear
EOS, M1 neutrino transport and a nuclear reaction network. Flares with $B = 10^{15}$--$10^{16.5}\,$G unbind
$10^{-10}$--$5\times10^{-8}\,M_\odot$ of the outermost crust, and the ejected mass grows roughly in proportion to the
flare pressure, with about $10^{-3}$ of the flare energy ending up as kinetic energy of the ejecta. Neutrinos matter
only in the strongest flares. At $10^{16.5}\,$G the flare layer radiates about a tenth of its energy in neutrinos
while it still drives the shock, and the ejecta mass is halved. The ejecta keep the electron fraction of the crust,
$\Ye \approx 0.44$--$0.47$, because the weak interactions are fast enough to change it only above
$\sim10^{16}\,$G, and even there they change it by only about $0.01-0.02$.

The ejecta leave at about half the speed of light with entropies of a few hundred $k_B$ per baryon. They expand too fast to build seed nuclei and freeze out as $\alpha$ particles and free neutrons, and nuclei with $A \geq 90$ make up a mass of at most $3.5\times10^{-11}\,M_\odot$, even in the strongest flare. This is three to four orders of magnitude below the analytic model of \citet{Patel:2025tse}, whose ejecta reach deeper, colder crust that our flares do not unbind. How far these numbers apply to real flares depends on our assumptions, and what they imply can be tested against the 2004 flare.

\emph{Upper limits for realistic flares.} Our models share with earlier work the assumptions of spherical symmetry, a flare layer that covers the whole star
and releases its energy at once, and no magnetic field in the crust. We expect most of these to overestimate the ejection. A real flare releases its energy over part of the surface, where the layer can also expand sideways and press on the crust for a shorter time, and where the bound crust falls back less coherently than in one dimension,
so the later shocks it drives are likely weaker. The 2004 outflow was likely one-sided \citep{Taylor:2005iv}. If the crust carries a field comparable to the one that powers the flare, the gas we eject has a magnetization $\sigma \approx 6$--$70$. Matter can then leave freely only along open field lines, which our models represent, while
a tangential field holds the crust down and a strongly magnetized shock converts little of its energy into heat \citep{1984ApJ...283..694K}. The field also changes the composition of the outer crust \citep{Servais:2026cng}. If the flare
layer itself is magnetically dominated, as the trapped fireball of \citet{Chatterjee:2026tdp}, its plasma pressure is
well below the $B^2/8\pi$ we assume. Only a slower release of the energy has an effect we cannot predict. Wider
layers change our ejecta mass by less than a factor of two, but a release over times longer than the expansion of
the shocked crust is outside our models. Since the heavy-element fraction falls by two to three orders of magnitude
per half decade in $B$, any of these effects would reduce the $r$-process yield even more than the mass. We
therefore regard our results as upper limits. Testing them requires multidimensional general-relativistic
magnetohydrodynamics with a localized energy release, open and closed field lines and a crust resolved to about a
meter, which local simulations of a patch of the surface could provide.

\emph{Radio afterglow.} For $B = 10^{15.5}$--$10^{16}\,$G our ejecta have $2\times10^{-9}$--$2\times10^{-8}\,M_\odot$
at $\approx0.45\,c$, enough for the radio afterglow of the 2004 flare, which requires $\gtrsim1.6\times10^{-9}\,M_\odot$,
and at the low end of the masses favored by \citet{Gelfand:2005wi} and \citet{Granot:2005vq}. These flares, however,
release $2\times 10^{48}$--$2\times 10^{49}\,$erg if the flare layer covers the whole star, $60$--$600$ times the
${\sim}3\times10^{46}\,$erg radiated in the prompt $\gamma$-ray spike, which measures the energy released by the flare.
A flare that covers only part of the surface releases less energy, but it also ejects less mass, because in our
models the ejecta mass is a nearly fixed fraction of the flare energy, $\Mej \approx (1$--$2)\times10^{-3}\,E_{\rm
fb}/c^2$. A flare with the energy of the spike therefore ejects only $\approx(2$--$3.5)\times10^{-11}\,M_\odot$, 50-100 times less than the afterglow requires, however large the area it covers. In the picture of
\citet{Patel:2025frn}, the ${\sim}10^{-6}\,M_\odot$ that produce both the afterglow and the delayed MeV emission carry a kinetic energy comparable to that of the spike, so the flare would have to give most of its energy to the crust,
while our flares give it about $10^{-3}$ of theirs. Unless the flare released much more energy than its $\gamma$-rays
show, or a real geometry ejects the crust far more efficiently than our one-dimensional models, the crust shock
alone cannot supply the baryons of the 2004 afterglow.

\emph{$r$-process and the MeV emission of the 2004 flare.} If giant flares eject crust as in our models, they are not a significant source of $r$-process elements. The
Galactic production rate of ${\sim}10^{-9}\,M_\odot\,{\rm yr^{-1}}$ that C24 estimated by assuming that all the ejecta
become $r$-process material drops by three orders of magnitude or more. The same applies to predictions that rest on
a large heavy-element fraction in the ejecta, such as MeV $\gamma$-ray lines \citep{Qiumu:2026xul}, ultraheavy cosmic
rays \citep{Patel:2026ynu} and the kilonovae of repeated flares in neutron-star--white-dwarf mergers
\citep{Zhong:2026loz}. The delayed MeV emission of the 2004 flare was interpreted by \citet{Patel:2025frn} as the
decay of $\approx7\times10^{-7}\,M_\odot$ of freshly made $r$-process nuclei. Our largest yield of nuclei heavier
than $\alpha$ particles is $\approx 8\times10^{-11}\,M_\odot$ (B16.5), four orders of magnitude less.

\emph{A neutron-powered nova brevis.} The ejecta can still power a short optical and ultraviolet transient through the $\beta$-decay of their free
neutrons, which make up $7$--$10$ percent of their mass against at most $8\times10^{-4}$ in heavy nuclei. With a mean
life of $\approx 900\,$s the neutrons heat the ejecta at
$\dot\epsilon \approx 3\times10^{14}\,X_n\,e^{-t/900\,{\rm s}}\,{\rm erg\,g^{-1}\,s^{-1}}$
\citep{2005astro.ph.10256K,Metzger:2014yda}, assuming that the decay electron thermalizes. For B16.0 this gives $\approx10^{39}\,{\rm erg\,s^{-1}}$, comparable to the
$r$-process-powered transient that C24 estimated. The ejecta are mostly helium and hydrogen, with an opacity close
to that of electron scattering, so the transient peaks after only $\approx40$--$55\,$s and is hotter than an
$r$-process-powered one. This means that for giant flares considered in this work, neutron decay would be the main power source of an optical transient.

\begin{acknowledgments}
TP acknowledges support from NSF Grant PHY-2020275 (Network for Neutrinos, Nuclear Astrophysics, and Symmetries (N3AS)) and DOE award DE-SC0004658.
DR acknowledges support from NASA through Grant No. 80NSSC25K7213, from the Department of Energy, Office of Science, under Awards Numbers DE-SC0024388 and DE-SC0026818, and from the National Science Foundation under Grants No. PHY‐2407681, PHY-2512802, PHY‐2608045, and PHY‐2621752.
JF acknowledges support from the National Aeronautics and Space Administration under Grant No. 80NSSC25K213.
YQ acknowledges support from the
Department of Energy, Office of Science, Division of Nuclear Physics, under
Award Numbers DE-SC0024388 and DE-SC0026818.
\end{acknowledgments}

\software{\athenak\ \citep{stone2024athenakperformanceportableversionathena,Zhu:2024utz,Fields:2024pob}, Kokkos \citep{9485033},
PrimitiveSolver \citep{Cook:2023bag}, \nurates\ \citep{Chiesa:2024lnu}, SkyNet \citep{Lippuner:2017tyn},
NumPy \citep{Harris:2020xlr}, SciPy \citep{2020NaMet..17..261V}, Matplotlib \citep{4160265}, h5py.}

\appendix

\section{Low-density extension of the EOS table}\label{app:eos}

The HS(DD2) table ends at $n_{\rm min} = 10^{-12}\,{\rm fm^{-3}}$ ($1.7\times10^{3}\,\gcc$). Below it we add 150
rows with the spacing of the table, $\Delta\log_{10} n_b = 0.04$, down to $10^{-18}\,{\rm fm^{-3}}$ ($1.7\times10^{-3}\,\gcc$), and keep its $\Ye$ and $T$ axes. At these densities matter is an ideal gas of nuclei, nucleons and electrons, together with photons and thermal $e^\pm$ pairs. The pressure of the gas is proportional to
$n_b$, while that of the photons and pairs depends only on $T$ and is already 30 times larger at $n_{\rm min}$ and $T = 0.1\,\MeV$, and more at higher $T$. In every column of fixed $\Ye$ and $T$ we therefore split each quantity at $n_{\rm min}$ into a gas part and a part that does not depend on density, which we call radiation, and continue
each part to lower density with the dependence it has for an ideal gas or for radiation. All constants are read off the table at its edge, so that the extension joins the table in every column without a step. 
The pressure is
\begin{equation}
  P(n_b, \Ye, T) = (\Gamma - 1)\,\beta\,n_b\,T + P_{\rm rad}(\Ye, T), \qquad \Gamma = 5/3 ,
\end{equation}
where $\beta$ follows from the pressures at the two lowest densities of the table and $P_{\rm rad}$ is the pressure
at $n_{\rm min}$ minus the gas term. The entropy of the radiation is obtained in the same way, as the tabulated entropy
at $n_{\rm min}$ minus that of the gas (Sackur--Tetrode, for $(\Gamma - 1)\beta$ particles per baryon), and its energy
density is $u_{\rm rad} = T s_{\rm rad} - P_{\rm rad}$. Below $n_{\rm min}$ the pressure thus becomes independent of
density, and the entropy and energy per baryon grow as $1/n_b$. The sound speed is recomputed from $P$, $s$ and $e$
and changes smoothly across $n_{\rm min}$. The particle fractions are kept at their values at $n_{\rm min}$. The further dissociation of nuclei that NSE would give at lower density involves at most ${\approx}8\,\MeV$ per nucleon, small against the radiation energy of more than $300\,\MeV$ per baryon where it occurs. The chemical potentials, which enter the neutrino rates, are continued from $n_{\rm min}$ with the density dependence of a dilute gas, that of the electrons from a nondegenerate $e^\pm$ gas in pair equilibrium, $\mu_e = T\,{\rm arsinh}[\Ye n_b/2n_\pm(T)] + {\rm const}$, with $n_\pm(T)$ the density scale of the thermal pairs.

\section{Numerical details}\label{app:numerics}

\paragraph{First-order flux correction.}
Our first-order flux correction differs from that of \citet{Fields:2024pob} in two points. The bounds of the
discrete maximum principle are $[\min/M,\ M\max]$ for positive and $[M\min,\ \max/M]$ for negative neighbor values,
with $M = 1.2$, because the conserved energy $\tau$ is negative in the outer crust, where the nuclei are bound. The
test for $\Ye$ is applied to $\Ye$ itself, with an absolute tolerance of $10^{-6}$ and the range of the EOS table,
instead of to the conserved $D\Ye$, and flagged cells receive first-order fluxes of $\Ye$ as well. With the HLLD
solver the first-order fluxes are computed with HLLE.

\paragraph{Well-balanced reconstruction.}
The well-balanced scheme is a relativistic extension of the method of \citet{2016A&A...587A..94K}, which is described in more
detail in Fields \& Dittmann (in preparation). Briefly, the method integrates the local state assuming hydrostatic
equilibrium to compute an equilibrium pressure, then uses piecewise-equilibrium reconstruction on the deviation from equilibrium
to compute the pressure at the left and right interfaces needed for the Riemann solver. The scheme employed here is an earlier
version of this method and therefore differs slightly from the algorithm described there in that
the perturbation is reconstructed only using piecewise-linear reconstruction with a van Leer slope limiter \citep{VANLEER1974361}.
Where the equilibrium integration would produce
a negative pressure, the well-balanced scheme is disabled and the pressure uses the standard WENO-Z reconstruction scheme of
the other primitive variables. This scheme can only accurately preserve equilibria if the Riemann solver preserves contact
discontinuities; this motivates our decision to employ an HLLD Riemann solver despite the system being unmagnetized. 

\bibliographystyle{aasjournal}
\bibliography{references}

\end{document}